\documentclass[lettersize,journal]{IEEEtran}

\makeatletter
\def\ps@headings{%
\def\@oddhead{\mbox{}\scriptsize\rightmark \hfil \thepage}%
\def\@evenhead{\scriptsize\thepage \hfil \leftmark\mbox{}}%
\def\@oddfoot{}%
\def\@evenfoot{}}
\makeatother 
\IEEEoverridecommandlockouts
\usepackage{bbm}
\usepackage{amsfonts}
\usepackage[dvips]{graphicx}
\usepackage{times}
\usepackage{cite}
\usepackage{amsmath}
\usepackage{array}
\usepackage{amssymb}

\usepackage{stfloats}
\usepackage{graphicx}
\usepackage{footnote}
\usepackage{booktabs}
\usepackage{array}
\usepackage{algorithm}
\usepackage{subeqnarray}
\usepackage{cases}
\usepackage{threeparttable}
\usepackage{color}
\usepackage{hyperref}
\usepackage{epstopdf}
\usepackage{algpseudocode}
\usepackage{bm}
\usepackage{multirow}
\usepackage[labelformat=simple]{subcaption}
\usepackage{adjustbox}

\usepackage{enumitem}
\usepackage{algpseudocode}
\usepackage{algorithm}
\usepackage{xcolor,etoolbox}
\usepackage{placeins}
\usepackage{tikz}

\usepackage{array}
\newcolumntype{C}[1]{>{\centering\arraybackslash}p{#1}}

\let\mybibitem\bibitem
\renewcommand{\bibitem}[1]{%
\ifstrequal{#1}{8778671}{\color{black}\mybibitem{#1}}
{\ifstrequal{#1}{wang2018preempt}{\color{black}\mybibitem{#1}}
{\ifstrequal{#1}{kavitha2018controlling}{\color{black}\mybibitem{#1}}
{\color{black}\mybibitem{#1}}}}%
}

\allowdisplaybreaks

\begin{document}

\title{
Radio Map Updating from Streaming Spectrum Measurements via Memory-Based Online Gaussian Processes
}
\author{Yuanyuan~Deng,
        Bo~Zhou,~\IEEEmembership{Member,~IEEE,}
        Tian~Chen,
        Shijian~Gao,~\IEEEmembership{Member,~IEEE,}
        Jia~Yan,~\IEEEmembership{Member,~IEEE,}
        Lantu~Guo,
        Qiuming~Zhu,~\IEEEmembership{Senior Member,~IEEE,}
        Qihui~Wu,~\IEEEmembership{Fellow,~IEEE,}
\thanks{This research was supported by the National Natural Science Foundation of China under Grants 62231015 and 62571237.
A preliminary version of this work has been presented at IEEE International Workshop on Machine Learning for Signal Processing, Aug./Sep., 2025, Istanbul, Turkey \cite{deng2025}.

Y.~Deng, B.~Zhou, T.~Chen, Q.~Zhu and Q.~Wu are with the College of Electronic and Information Engineering, Nanjing University of Aeronautics and Astronautics, Nanjing, China.
S.~Gao is with the Internet of Things Thrust, The Hong Kong University of Science and Technology, Guangzhou, China.
J.~Yan is with the Intelligent Transportation Thrust, The Hong Kong University of Science and Technology, Guangzhou, China.
L~ Guo is with China Research Institute of Radiowave Propagation,
Qingdao 266107, China.
Email: \{yydeng0601, b.zhou, chentian123, zhuqiuming, wuqihui\}@nuaa.edu.cn, \{shijiangao, jasonjiayan\}@hkust-gz.edu.cn, guolantu@163.com.
}
}

\maketitle

\begin{abstract}
Radio maps, which estimate spatial radio-frequency characteristics from spectrum measurements, are essential for applications such as spectrum management and network planning.
With the continuous arrival of spectrum measurements, 
conventional batch processing methods for radio map reconstruction become computationally prohibitive, as they require reprocessing all accumulated measurements for each radio map update.
To address this, we propose a memory-based online sparse variational Gaussian process (M-OSVGP) method that  efficiently updates radio maps from streaming spectrum measurements.
Our method employs sparse variational inference and updates the posterior online by minimizing a hybrid objective that integrates newly received measurements and a memory subset of previous ones to mitigate catastrophic forgetting.
To further improve posterior approximation as measurements accumulate over spatially diverse regions, we extend M-OSVGP with a grid-assisted online inducing point selection (GOIPS) algorithm. GOIPS dynamically adapts the number and locations of inducing points based on measurement density and spatial correlation, providing a more informative inducing set while maintaining computational efficiency.
Extensive simulations demonstrate the effectiveness of our proposed methods in  reconstruction accuracy, computational efficiency, and uncertainty quantification, compared to existing batch and online baselines across various scenarios.

\end{abstract} 

\begin{IEEEkeywords}
Radio map, online Gaussian process, sparse approximation, variational inference, inducing point selection.
\end{IEEEkeywords}

\section{Introduction}
Radio maps provide spatial representations of the characteristics of the radio-frequency spectrum that capture information about spectral activities and propagation channels within a geographic area \cite{Bi2019,10430216}. 
They are essential tools for diverse applications including wireless spectrum allocation \cite{Romero2022}, network resource management \cite{Zhang2020,11592670,11016100}, anomaly detection and recognition \cite{10254521,11442296}, and source localization \cite{Chenjunting2019, Yapar2023,He2023}.
Constructing a radio map involves estimating the signal distribution across a target area using spectrum measurements from spatially distributed sensors. 
In practice, spectrum measurements often arrive continuously in batches at the monitoring center, for instance, from sensors that are  newly deployed or relocated to previously unmonitored areas.
To maintain an accurate representation of the spectrum environment, it is  imperative to  process these streaming measurements efficiently and  update the radio map dynamically over time.

Current methods for radio map construction can be broadly categorized into model-driven and data-driven approaches. 
Model-driven methods leverage prior knowledge of emitters or  radio propagation models, such as empirical path-loss models\cite{Iskander2002}, dominant path models\cite{Wahl2005}, and ray-tracing models\cite{Suga2021}, to infer the spectrum distribution.
These methods typically require detailed environmental modeling and extensive ray-level computation, making them computationally expensive, and their accuracy may degrade significantly when the assumed propagation models do not align with real-world environments.
In contrast, data-driven methods exploit statistical properties to construct radio maps by interpolating signal measurements, without requiring detailed environmental or emitter-specific information. 
Representative methods include interpolation methods \cite{Cover1967, Chiles2012, Sato2017}, Gaussian processes (GP)-based methods \cite{Polyzos2024}, matrix completion \cite{Sun2022}, and tensor completion \cite{Zhang2020a, Sunhao2024}, which can offer greater flexibility across different deployment scenarios.
Specifically, interpolation methods estimate the received signal strength (RSS) value at each map point as a weighted combination of nearby measurements, where techniques such as k-nearest neighbors (KNN) \cite{Cover1967}, inverse distance weighting (IDW) \cite{Chiles2012}, and Kriging \cite{Sato2017} provide different strategies for determining the weights.
Matrix or tensor completion methods formulate the task as a completion problem and estimate missing values by exploiting the low-rank structure of the spectrum measurement matrix.
More recently, deep learning (DL) has been  applied  for radio map reconstruction \cite{Levie2021, Teganya2021, Li2024, 11066175,11455177}, as DL can learn complex, non-linear spatial relationships in measurements to capture multi-path and shadowing effects.
However, these methods typically require large labeled datasets for training, which are often impractical to acquire, and their ``black-box'' nature  further complicates their deployment in real-world wireless systems.

It is worth noting that most existing radio map construction methods follow a batch processing paradigm, where the radio map is constructed from scratch by reprocessing all received measurements upon the arrival of new spectrum measurements. 
This results in high computational overhead and  limited scalability, rendering such batch processing-based methods unsuitable for scenarios with continuous measurement streams.
To address these limitations, several works have explored online approaches for radio map construction.
For instance, reference \cite{Huang2019} studies WiFi fingerprint-based localization with noisy location labels through online radio map constructions with crowdsourced RSS data. 
The work in \cite{Zhao2021} proposes a generative adversarial network (GAN)-based method, in which new signal maps are generated using a pre-trained and fixed GAN generator once actively collected crowdsourced data is available.
For cellular coverage maps, the work in \cite{Kasparick2015} proposes two kernel-based online algorithms that leverage user trajectory information to assign larger weights to measurements near the user of interest's trajectory, while \cite{Nikbakht2018} introduces a dual-kernel online method that models the direct current (DC) and varying components of channel losses separately.
Despite these efforts, existing online methods still exhibit important limitations.
Note that the method in \cite{Zhao2021} does not adapt the GAN generator to newly arrived measurements, while the approaches in \cite{Kasparick2015} and \cite{Nikbakht2018} rely on specific prior information such as user trajectory information or pre-defined channel loss decomposition models, which may not be available in general scenarios.
Therefore, it remains a key challenge to design an online radio map construction framework that can efficiently and adaptively update the map from streaming measurements without relying on restrictive prior assumptions.

In this paper, we propose a memory-based online sparse variational Gaussian process (M-OSVGP) method  for efficient and adaptive radio map reconstruction from streaming spectrum measurements.
Specifically, built upon sparse variational inference, M-OSVGP initializes the inducing points by randomly selecting a subset of spectrum measurements, and then refines these locations using gradient descent. 
To online update the posterior, our method minimizes a hybrid objective function that integrates newly received measurements with a memory subset of previous ones, which helps mitigate catastrophic forgetting and maintain inference accuracy.  
Since sparse variational inference relies on inducing points to approximate the GP posterior, the quality of the inducing point set directly affects the reconstruction accuracy.
However, as measurements may arrive from spatially diverse regions over time, randomly initialized inducing points may fail to capture the spatial characteristics of the accumulated measurements, resulting in an inaccurate posterior approximation.
To address this issue, we further extend M-OSVGP with a grid-assisted online inducing point selection (GOIPS) algorithm, referred to as M-OSVGP-GOIPS. 
GOIPS adaptively updates both the locations and number of inducing points based on spatial correlation patterns and measurement density, ensuring that the inducing set captures the characteristics of the current environment effectively and provides a better starting point for subsequent gradient-based refinement.
Finally, we evaluate the effectiveness and robustness of our proposed methods, M-OSVGP and M-OSVGP-GOIPS, by comparing them with batch processing methods, including KNN, IDW, and sparse variational GP (SVGP) \cite{Titsias2009}, as well as online processing methods, including streaming SVGP (SSVGP) \cite{Bui2017} and memory-based dual GP (Memory-DGP) \cite{Chang2023}. Simulation results reveal that our methods achieve superior performance in both reconstruction accuracy and computational efficiency across different scenarios.

In summary, our key contributions include:
\begin{itemize}
\item[$\bullet$] We propose a novel online  framework for radio map reconstruction that efficiently processes streaming spectrum measurements. 
Our method can efficiently update the radio map while mitigating catastrophic forgetting, by simultaneously learning from new measurements and retaining knowledge from a representative memory of previous measurements.
\item[$\bullet$] We introduce the GOIPS algorithm to adaptively manage the inducing point set based on the spatial correlation and measurement density of the received measurements. GOIPS selects informative locations and adjusts the number of inducing points to improve posterior approximation accuracy and reduce computational cost.
\item[$\bullet$] We conduct extensive simulations under various scenarios, demonstrating that our proposed methods consistently outperform existing batch and online baselines in terms of reconstruction accuracy and  computational cost.  
Furthermore, we show that GOIPS can be integrated as a standalone module to enhance other inducing point-based GP methods.
\end{itemize}

The remainder of this paper is organized as follows. 
Section II introduces the system model for radio map updating. 
Section III presents GP-based radio map reconstruction methods in the batch processing mode.
In Section IV, we propose the online method, M-OSVGP, and its extension with the GOIPS algorithm for adaptive inducing point management.
Section V provides comprehensive simulation results and performance analysis.
Finally, Section VI concludes the paper.

\begin{figure}[!t]
    \centering
    \includegraphics[width=0.5\textwidth]{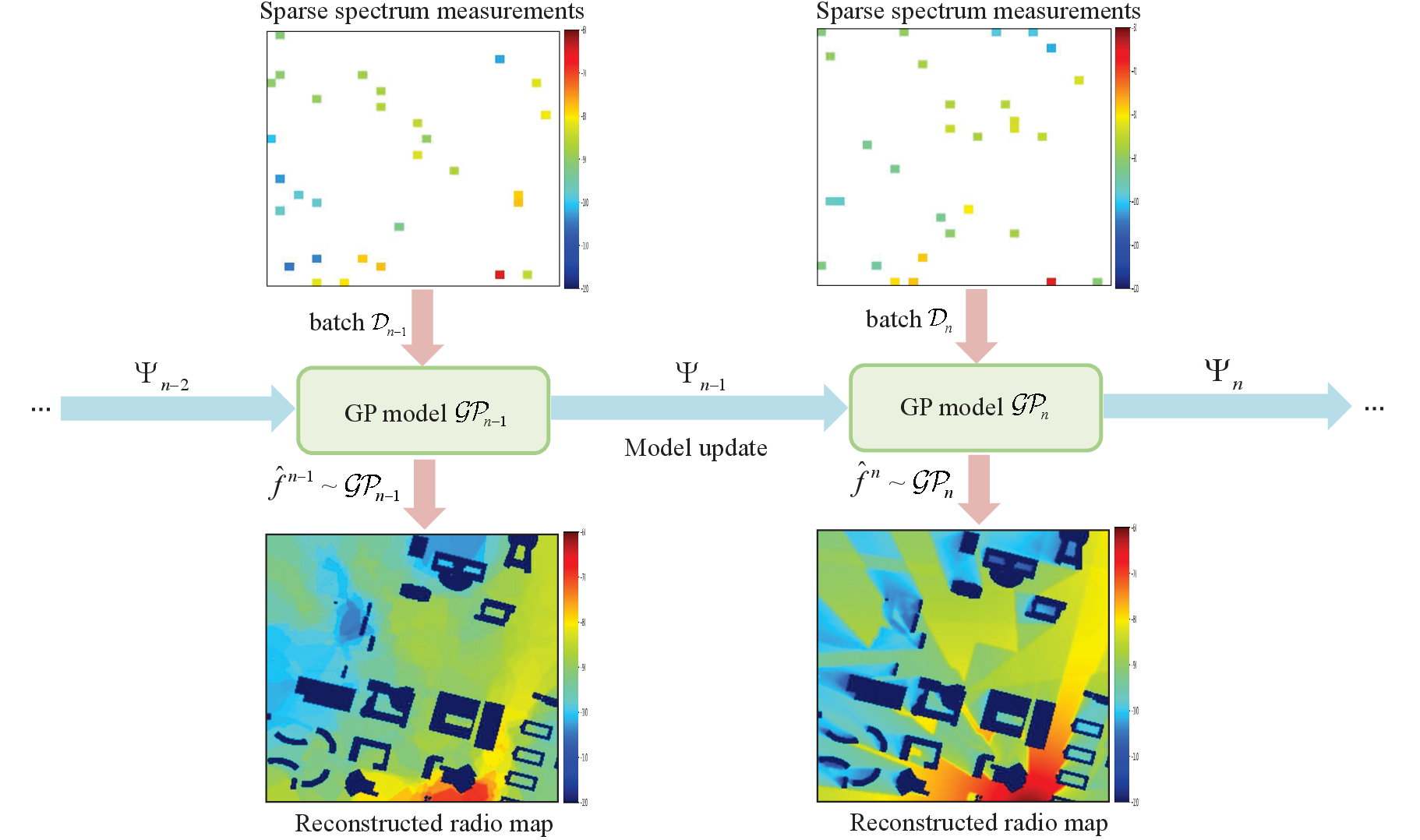}  
    \caption{The illustration of online radio map updating, where $\Psi_{n}=\{\mathbf{m}_{\mathbf{u}_n}, \mathbf{S}_{\mathbf{u}_n}, \mathbf{Z}_n, \theta_n\}$ are the parameters of the GP model in the $n$-th update.}
    \label{fig: system model}
\end{figure}

\section{System Model}
Consider a wireless coverage scenario over a geographic region of interest $\mathcal{X} \subset \mathbb{R}^2$, where distributed sensors continuously report the RSS measurements with the corresponding location information to the monitoring center for radio map reconstruction. 
The region $\mathcal{X}$ is discretized into a rectangular grid $\mathcal{G}$ of dimensions $H \times W$, with uniform grid sizes in both directions. 
The monitoring center updates the radio map whenever it has accumulated a batch containing a certain number of new measurements.
Let $\mathcal{D}_n = \{\mathbf{X}_n, \boldsymbol{\psi}_n\}$ denote the $n$-th batch of the received  measurements, where $\mathbf{X}_n=[\boldsymbol{x}_{n, 1}, \boldsymbol{x}_{n, 2},...,\boldsymbol{x}_{n, N_n}]$ and $\boldsymbol{\psi}_n=[\psi_{n, 1}, \psi_{n, 2}, ...,\psi_{n, N_n}]$. 
Here, $N_n$ is the number of measurements in the $n$-th batch and $\psi_{n, i}$ denotes the RSS value at location $\boldsymbol{x}_{n, i}$. 
The cumulative dataset up to the $n$-th batch, denoted as $\mathcal{D}_{1:n}=\bigcup_{j=1}^n\mathcal{D}_j$, can be expressed as $\{\mathbf{X}_{1:n}, \boldsymbol{\psi}_{1:n}\}$, where $\mathbf{X}_{1:n}=[\mathbf{X}_1, \mathbf{X}_2,...,\mathbf{X}_n]$ and $\boldsymbol{\psi}_{1:n}=[\boldsymbol{\psi}_1, \boldsymbol{\psi}_2, ..., \boldsymbol{\psi}_n]$ represent all collected locations and their corresponding RSS values.
Let $\mathbf{X}_n^*$ denote the location set of grid points in $\mathcal{G}$ that remain unmeasured after receiving the $n$-th batch (referred to as non-measured points), and let $\boldsymbol{\psi}_n^*$ denote the RSS values at these points, which we aim to estimate to reconstruct the radio map.

For radio map reconstruction, the RSS measurement at each location $\boldsymbol{x}$ can be modeled as \cite{Polyzos2024,Kasparick2015, Nikbakht2018}:  
  \begin{align}\label{eqn:RSS model}
\psi = f(\boldsymbol{x})+\epsilon,
 \end{align}
where $f(\boldsymbol{x})$ is the true RSS value at location $\boldsymbol{x}$, modeled as an unknown latent function $f(\cdot): \mathbb{R}^2 \to \mathbb{R}$, and $\epsilon$ denotes zero-mean Gaussian measurement noise with variance $\sigma_{\psi}^2$.
We model the latent function $f$ as a GP, defined as: 
\begin{align}\label{function_GP}
f \sim \mathcal{GP}(m(\boldsymbol{x}), k(\boldsymbol{x}, \boldsymbol{x}')), 
\end{align}
where $m(\boldsymbol{x})$ denotes the mean function, representing the prior expectation of the signal strength at location $\boldsymbol{x}$ and $k(\boldsymbol{x}, \boldsymbol{x}')$ is the kernel function capturing the similarity between the RSS values at locations $\boldsymbol{x}$ and $\boldsymbol{x}'$. 
Typical kernel functions include the Radial Basis Function (RBF), linear, polynomial, and Mat\'{e}rn kernels \cite{Williams2006}.

Our objective is to estimate the latent function as $\hat{f}^n$ upon receiving each new batch $\mathcal{D}_n$, thereby infer the missing RSS values $\boldsymbol{\psi}_n^*$ at non-measured points $\mathbf{X}_n^*$. 
This task can be solved using either batch or online processing methods.
Under batch processing, $\hat{f}^n$ is estimated directly using all the accumulated measurements $\mathcal{D}_{1:n}$. 
In contrast, under online processing, after receiving a new batch $\mathcal{D}_n$, $\hat{f}^n$ is dynamically updated by refining the previous estimate $\hat{f}^{n-1}$ for the $(n-1)$-th batch. 

In Section III, we will first describe how GP and its sparse variant, SVGP, can be applied for radio map reconstruction in the batch processing mode. Then, in Section IV, we will propose  our online radio map updating method, M-OSVGP, which builds on the SSVGP framework and incorporates a memory subset of previous measurements to mitigate catastrophic forgetting.
We will further introduce the GOIPS algorithm to dynamically select inducing points by leveraging spatial information.


\section{GP-based Radio Map Reconstruction Under Batch Processing}
In this section, we present GP-based methods for radio map reconstruction under batch processing, where all accumulated measurements are reprocessed at each update step.
We begin with standard GP, which offers accurate spatial predictions but suffers from high computational cost. 
To address this limitation, we then describe the SVGP framework \cite{Titsias2009}, which uses a small set of inducing points to achieve scalable inference \footnote{In Section 3,  we omit the batch index subscript $n$ for clarity, as all operations are performed on the entire accumulated dataset.}.  

\subsection{Standard GP for Radio Map Reconstruction}
GPs provide a foundational framework for radio map reconstruction due to their effectiveness in spatial interpolation and function estimation tasks \cite{Williams2006,Chen2022}.
Given the accumulated measurements $\mathcal{D}=\{\mathbf{X},\boldsymbol{\psi}\}$ till the $n$-th batch, we aim to estimate RSS values $\boldsymbol{\psi}^*$ at locations $\mathbf{X}^*$. Since any finite set of samples drawn from a GP follows a joint Gaussian distribution \cite{Williams2006}, the joint distribution of RSS measurements $\boldsymbol{\psi}$ and the target RSS values $\boldsymbol{\psi}^*$ can be expressed as
\begin{align}
\begin{bmatrix} \boldsymbol{\psi} \\ \boldsymbol{\psi}^* \end{bmatrix} \sim \mathcal{N}
\left(\begin{bmatrix} \boldsymbol{u} \\ \boldsymbol{u}^* \end{bmatrix},
\begin{bmatrix} \mathbf{K}_{\boldsymbol{\psi\psi}}+\sigma_{\psi}^2\mathbf{I} & \mathbf{K}_{\boldsymbol{\psi}\boldsymbol{\psi}^*} \\ \mathbf{K}_{\boldsymbol{\psi}^*\boldsymbol{\psi}} & \mathbf{K}_{\boldsymbol{\psi}^*\boldsymbol{\psi}^*} \end{bmatrix}\right).
 \end{align}
Here, $\boldsymbol{u}$ and $\boldsymbol{u}^*$ denote the prior mean vectors for the measured and target locations, respectively (typically assumed to be zero or constant), and $\mathbf{I}$ is an identity matrix.
$\mathbf{K}_{\mathbf{\boldsymbol{\psi}\boldsymbol{\psi}}}$ is the symmetric positive definite covariance matrix of the measurement locations $\mathbf{X}$ with its $(p, q)$-th element being $[\mathbf{K}_{\mathbf{\boldsymbol{\psi}\boldsymbol{\psi}}}]_{p,q} = k(\boldsymbol{x}_p, \boldsymbol{x}_q)$, $\mathbf{K}_{\mathbf{\boldsymbol{\psi}\boldsymbol{\psi}^*}}$, $\mathbf{K}_{\mathbf{\boldsymbol{\psi}^*\boldsymbol{\psi}}}$ is the cross-covariance matrix between $\mathbf{X}$ and $\mathbf{X}^*$, and $\mathbf{K}_{\mathbf{\boldsymbol{\psi}^*\boldsymbol{\psi}^*}}$ is the covariance matrix of the non-measured locations $\mathbf{X}^*$.
By applying the standard conditioning rules for multivariate Gaussian distributions, the posterior distribution over $\boldsymbol{\psi}^*$ becomes
\begin{align}
p(\boldsymbol{\psi}^*|\boldsymbol{\psi},\mathbf{X},\mathbf{X}^*) = \mathcal{N}(\boldsymbol{\mu}_{\text{GP}}^*,\mathbf{\Sigma}_{\text{GP}}^*),
\end{align}
where
\begin{align}
&\boldsymbol{\mu}_{\text{GP}}^*=\boldsymbol{u}^*+\mathbf{K}_{\boldsymbol{\psi}^*\boldsymbol{\psi}}\left(\mathbf{K}_{\boldsymbol{\psi\psi}}+\sigma_{\psi}^2\mathbf{I}\right)^{-1}\left(\boldsymbol{\psi}-\boldsymbol{u}\right),\label{eq:prediction}\\
&\mathbf{\Sigma}_{\text{GP}}^*=\mathbf{K}_{\boldsymbol{\psi}^*\boldsymbol{\psi}^*}-\mathbf{K}_{\boldsymbol{\psi}^*\boldsymbol{\psi}}\left(\mathbf{K}_{\boldsymbol{\psi\psi}}+\sigma_{\psi}^2\mathbf{I}\right)^{-1}\mathbf{K}_{\boldsymbol{\psi\psi}^*}.
\end{align}
Thus, the posterior mean $\boldsymbol{\mu}_{\text{GP}}^*$ provides the RSS estimates at the non-measured locations $\mathbf{X}^*$, while the diagonal elements of the posterior covariance $\mathbf{\Sigma}_{\text{GP}}^*$ quantify the corresponding estimation uncertainty.

The GPs are non-parametric models that do not specify an explicit parametric form for the function $f$. 
Instead, a GP represents a distribution over functions by defining the covariance between function values at different inputs through a kernel function $k(\boldsymbol{x},\boldsymbol{x}')$. 
This kernel encodes similarity between input locations and is parameterized by hyperparameters  that control properties such as smoothness and amplitude.
A commonly used kernel is the RBF kernel, given by
\begin{align*}
    k(\boldsymbol{x},\boldsymbol{x}') = \sigma_f^2 \exp\left(-\frac{|\boldsymbol{x}-\boldsymbol{x}'|^2}{2\ell^2}\right),
\end{align*}
where the signal variance $\sigma_f^2$ determines the vertical scale of the function and the length-scale $\ell$ controls how quickly the correlation between points decays with distance.
The Mat\'{e}rn kernel generalizes the RBF kernel by introducing a smoothness parameter $\nu$, given by
\begin{align*}
    k(\boldsymbol{x},\boldsymbol{x}') = \sigma_f^2\frac{2^{1 - \nu}}{\Gamma(\nu)} \left( \frac{\sqrt{2\nu} \, |\boldsymbol{x}-\boldsymbol{x}'|}{\ell} \right)^{\nu} K_{\nu} \left( \frac{\sqrt{2\nu} \, |\boldsymbol{x}-\boldsymbol{x}'|}{\ell} \right),
\end{align*}
where $K_{\nu}(\cdot)$ is the modified Bessel function and $\Gamma(\cdot)$ is the Gamma function.
The parameter $\nu$  directly controls the smoothness of the function: smaller values produce rougher functions, while larger values lead to smoother ones.

To effectively use a GP model, the hyper-parameters $\theta$, such as $\sigma_f^2 $ and $ \ell$ from the RBF kernel and the measurement noise variance $\sigma_\psi^2$, should be determined from the measurements.
This can be achieved by maximizing the log marginal likelihood of the measurements \cite{Williams2006}. 
The marginal likelihood $p(\boldsymbol{\psi}|\theta)$ is obtained by integrating out the latent function values $\mathbf{f}=f(\mathbf{X})$ , where the GP prior is $p(\mathbf{f}|\theta) =  \mathcal{N}(0,\mathbf{K}_{\boldsymbol{\psi\psi}})$ and the likelihood is $p(\boldsymbol{\psi}|\mathbf{f})= \mathcal{N}(\mathbf{f},\sigma_{\psi}^{2}\mathbf{I})$. This leads to
\begin{align}
    p(\boldsymbol{\psi}|\theta)&=\int p(\boldsymbol{\psi}|\mathbf{f})p(\mathbf{f}|\theta)d\mathbf{f}= \mathcal{N}(0, \mathbf{K}_{\boldsymbol{\psi\psi}}+\sigma_{\psi}^{2}I).
\end{align}
Taking the logarithm yields the log marginal likelihood:
\begin{align}\label{eq:NLML}
\log p(\boldsymbol{\psi}|\theta) 
&=-\frac{N}{2}\log2\pi-\frac{1}{2}\log\left|\mathbf{K}_{\boldsymbol{\psi\psi}}+\sigma_{\psi}^{2}\mathbf{I}\right| \nonumber\\
&-\frac{1}{2}\left(\boldsymbol{\psi}-\bm{\mu}\right)^\top\left(\mathbf{K}_{\boldsymbol{\psi\psi}}+\sigma_{\psi}^{2}\mathbf{I}\right)^{-1}\left(\boldsymbol{\psi}-\bm{\mu}\right).
\end{align}

Note that computing \eqref{eq:NLML} in standard GP-based radio map reconstruction requires inverting the covariance matrix $\mathbf{K}_{\boldsymbol{\psi\psi}}+\sigma_{\psi}^{2}\mathbf{I}$ corresponding to all received RSS values $\boldsymbol{\psi}$ up to the  $n$-th batch. 
Under batch processing, where all these measurements are processed jointly, the computational complexity grows as $\mathcal{O}((\sum_{j=1}^nN_j)^3)$, which could become computationally intractable for a large number of accumulated measurements.

\subsection{Sparse Variational GP for Radio Map Reconstruction}
To reduce the high computational cost of standard GP in radio map reconstruction, we adopt the SVGP framework \cite{Titsias2009}, which introduces a set of \textit{inducing points} to summarize the measurement data and applies variational approximation for scalable inference. 
Specifically, given the accumulated measurements $\mathcal{D}=\{\mathbf{X},\boldsymbol{\psi}\}$ till the $n$-th batch, a set of $M\ll\sum_{j=1}^nN_j$ inducing points $\{\mathbf{Z}, \mathbf{u}\}$ is introduced, where the inducing variables $\mathbf{u}=f(\mathbf{Z})$ represent the latent function values at the corresponding inducing locations $\mathbf{Z}=[\boldsymbol{z}_1, \boldsymbol{z}_2, ...,\boldsymbol{z}_M]$.
The inducing variables $\mathbf{u}$ are a subset of the latent function values $\mathbf{f}=\{\mathbf{u, f_{\ne \mathbf{u}}}\}$ at locations $\mathbf{X}$, with $\mathbf{f}_{\ne \mathbf{u}}$ denoting the remaining elements of $\mathbf{f}$ excluding $\mathbf{u}$. 
Under SVGP, the exact posterior $p(\mathbf{f}|\boldsymbol{\psi})$ is approximated as the variational distribution $q(\mathbf{f})=q(\mathbf{u})p(\mathbf{f}_{\ne \mathbf{u}}|\mathbf{u},\theta)$, where $q(\mathbf{u})=\mathcal{N}(\mathbf{m_u},\mathbf{S_u})$ and $\theta$ are the kernel hyperparameters \cite{Titsias2009}. 
The optimal variational parameters $\{\mathbf{m_u},\mathbf{S_u}\}$, the inducing locations $\mathbf{Z}$, and hyperparameters $\theta$ can be obtained by 
maximizing the evidence lower bound (ELBO) on the log marginal likelihood $\log p(\boldsymbol{\psi}|\theta)$. 
The ELBO, denoted as $\mathcal{L}_\text{batch}$, can be derived by applying Jensen's inequality,
\begin{align} \label{LML_SVGP}
    \log p(\boldsymbol{\psi}|\theta)&=\log \int p(\boldsymbol{\psi},\mathbf{f}|\theta)d\mathbf{f}\nonumber\\
    &=\log \int q(\mathbf{f})\frac{p(\boldsymbol{\psi},\mathbf{f}|\theta)}{q(\mathbf{f})}d\mathbf{f} \nonumber\\
    &\geq \int q(\mathbf{f}) \log \frac{p(\boldsymbol{\psi},\mathbf{f}|\theta)}{q(\mathbf{f})}d\mathbf{f} \triangleq \mathcal{L}_\text{batch}.
\end{align}
 Note that, the ELBO can also be expressed as the difference between the model log-marginal and the Kullback–Leibler (KL) divergence between the variational distribution and the true posterior, i.e., $\mathcal{L}_\text{batch}= \log p(\boldsymbol{\psi}|\theta)-\text{KL}[q(\mathbf{f})||p(\mathbf{f}|\boldsymbol{\psi},\theta)]$, where $\text{KL}[\cdot||\cdot]$ denotes the  KL divergence. 
Thus, maximizing the ELBO is equivalent to minimizing the KL divergence between the variational distribution $q(\mathbf{f})$ and the true posterior $p(\mathbf{f}|\boldsymbol{\psi},\theta)$, thereby driving the approximation closer to the exact posterior.
Substituting $q(\mathbf{f})=q(\mathbf{u})p(\mathbf{f}_{\ne \mathbf{u}}|\mathbf{u},\theta)$ into \eqref{LML_SVGP}, we can further decompose this lower bound into:
\begin{align}\label{SVGP ELBO}
    \mathcal{L}_\text{batch}=&
    \int q(\mathbf{u})p(\mathbf{f}_{\ne \mathbf{u}}|\mathbf{u},\theta) \log \frac{p(\boldsymbol{\psi} | \mathbf{f},\theta) p(\mathbf{f}_{\ne \mathbf{u}}|\mathbf{u},\theta)p(\mathbf{u}|\theta)}{q(\mathbf{u})p(\mathbf{f}_{\ne \mathbf{u}}|\mathbf{u},\theta)}d\mathbf{f}\nonumber\\
    =& \sum_{\psi_i\in \boldsymbol{\psi}}\mathbb{E}_{q_{\mathbf{u}}(f_i)}[\log p(\psi_i|f_i)] - \text{KL}[q(\mathbf{u}) || p(\mathbf{u}|\theta)],
\end{align}
where $f_i=f(\boldsymbol{x}_i)$ is the function value at the measured points $\boldsymbol{x}_i$ and $q_{\mathbf{u}}(f_i) = \int p(f_i|\mathbf{u}, \theta)q(\mathbf{u})d\mathbf{u}$ is the marginal variational distribution over $f_i$.

Finally,  with the variational parameters optimized by maximizing the ELBO in \eqref{SVGP ELBO}, we can approximate the posterior distribution of the RSS values $\boldsymbol{\psi}^*$ at unmeasured locations $\mathbf{X}^*$ as follows:
\begin{align}
p(\boldsymbol{\psi}^*|\boldsymbol{\psi}) \approx
q(\boldsymbol{\psi}^*)=  \mathcal{N}\left(\boldsymbol{\mu}_{\text{SVGP}}^*,\mathbf{\Sigma}_{\text{SVGP}}^*\right),
\end{align}
where 
\begin{align}
&\boldsymbol{\mu}_{\text{SVGP}}^*=\mathbf{K}_{\boldsymbol{\psi}^*\mathbf{u}}\mathbf{K}_{\mathbf{uu}}^{-1}\mathbf{m}_\mathbf{u},\label{eq:prediction_1}\\
&\mathbf{\Sigma}_{\text{SVGP}}^*=\mathbf{K}_{\boldsymbol{\psi}^*\boldsymbol{\psi}^*}-\mathbf{K}_{\mathbf{\boldsymbol{\psi}^*u}}\mathbf{K}_{\mathbf{uu}}^{-1}\left(\mathbf{K}_{\mathbf{uu}}-\mathbf{S}_\mathbf{u}\right)\mathbf{K}_{\mathbf{uu}}^{-1}\mathbf{K}_{\mathbf{u\boldsymbol{\psi}^*}}.\label{eq:prediction_2}
 \end{align}
Here, $\mathbf{K}_{\mathbf{\boldsymbol{\psi}^*u}}$ is the covariance matrix between $\mathbf{X}^*$ and $\mathbf{Z}$. 
The computational complexity of SVGP-based radio map reconstruction is primarily dominated by operations involving the covariance between all received measurements up to the $n$-th batch and the inducing points, resulting in a more scalable complexity of $\mathcal{O}((\sum_{j=1}^nN_j)M^2)$ \cite{Titsias2009}. 

\section{Proposed Radio Map Updating Method}
Although the SVGP method can reduce the computational complexity of standard GP, it still requires recomputing the posterior distribution from scratch using all accumulated measurements whenever a new batch arrives. This becomes increasingly computationally expensive as the total number of measurements $\sum_{j=1}^nN_j$ increases.
To address this, in this section, we propose an efficient online radio map updating method build upon SSVGP \cite{Bui2017}, which incrementally updates the posterior by incorporating arriving spectrum measurements.
Specifically, we first develop the M-OSVGP method to efficiently incorporate new measurements while using a memory mechanism to mitigate catastrophic forgetting.
We then  enhance M-OSVGP with the GOIPS algorithm to adaptively manage the inducing points, based on the spatial distribution of the streaming measurements.

\subsection{Online Sparse Variational GP} \label{Approximated Online Sparse GP}
The core idea of our online method is to incrementally update the variational posterior as a new batch of measurements arrives, rather than recomputing it from scratch. 
This is enabled by employing a set of inducing points to define a variational approximation that summarizes all received measurements. Upon receiving a new batch, we update the variational parameters by optimizing a hybrid objective function derived from an online variational inference scheme, avoiding the need to reprocess all previous measurements. 


\subsubsection{Incremental Posterior Update via Online Variational Inference}
Our goal is to efficiently estimate the posterior distribution  $p(\mathbf{f}| \boldsymbol{\psi}_{1:n})$ over the latent function $f$ after receiving a new batch of measurements $\boldsymbol{\psi}_{n}$, without reprocessing all previous measurements.
The exact posterior conditioned on all measurements up to batch $n$ can be expressed recursively using Bayes' rule:
\begin{align}
    p(\mathbf{f}| \boldsymbol{\psi}_{1:n})=\frac{1}{Z_{1}(\theta_n)}p(\mathbf{f}|\theta_n)p(\boldsymbol{\psi}_{1:n-1}|\mathbf{f})p(\boldsymbol{\psi}_n|\mathbf{f}),\label{posterior-n}
\end{align}
where $Z_{1}(\theta_{n})=p(\boldsymbol{\psi}_{1:n}|\theta_{n})$ is a normalization constant and $\theta_{n}$ represents the GP hyperparameters at batch $n$. 

To make this posterior computationally tractable, we follow the SVGP framework and introduce a variational approximation using inducing points.
Let $\mathbf{u}_{n}=f(\mathbf{Z}_{n})$ denote the latent function values at the inducing points $\mathbf{Z}_{n}$ after receiving the $n$-th batch of measurements.
The inducing points $\mathbf{Z}_{n}$ are updated when a new batch $\mathcal{D}_n$ arrives, where an initialization is constructed from both the previous inducing set and the new batch, and will be further refined during variational optimization.
We then define the variational distribution as $q^{n}(\mathbf{f})=p(\mathbf{f}_{\ne \mathbf{u}_{n}}|\mathbf{u}_{n},\theta_{n})q^{n}(\mathbf{u}_{n})$, where $q^{n}(\mathbf{u}_{n})=\mathcal{N}(\mathbf{m}_{\mathbf{u}_{n}},\mathbf{S}_{\mathbf{u}_{n}})$.
The optimal $q^n(\mathbf{f})$ can be determined by minimizing its KL divergence to the exact posterior, $\text{KL}[q^n(\mathbf{f}) || p(\mathbf{f}| \boldsymbol{\psi}_{1:n})]$.
To avoid recomputing the variational approximation from scratch whenever a new batch arrives, we incrementally update $q^{n}(\mathbf{f})$ to approximate $p(\mathbf{f}| \boldsymbol{\psi}_{1:n})$ using the previous variational approximation $q^{n-1}(\mathbf{f})$.
This distribution approximates the previous posterior $p(\mathbf{f}|\boldsymbol{\psi}_{1:n-1})$, which is given by:
\begin{align}
    q^{n-1}(\mathbf{f}) \approx p(\mathbf{f}|\boldsymbol{\psi}_{1:n-1})=\frac{1}{Z_{2}(\theta_{n-1})}p(\mathbf{f}|\theta_{n-1})p(\boldsymbol{\psi}_{1:n-1}|\mathbf{f}).\label{posterior-n-1}
\end{align}
Here $Z_{2}(\theta_{n-1})=p(\boldsymbol{\psi}_{1:n-1}|\theta_{n-1})$ is another normalization constant. 
From \eqref{posterior-n-1}, we obtain that $p(\boldsymbol{\psi}_{1:n-1}|\mathbf{f}) \approx \frac{Z_{2}(\theta_{n-1})q^{n-1}(\mathbf{f})} {p(\mathbf{f}|\theta_{n-1})}$ and then substitute it into \eqref{posterior-n} yielding:
\begin{align}\label{q^n(f)}
p(\mathbf{f}| \boldsymbol{\psi}_{1:n}) \approx \frac{Z_{2}(\theta_{n-1})}{Z_{1}(\theta_{n})}p(\mathbf{f}|\theta_{n})p(\boldsymbol{\psi}_{n}| \mathbf{f})\frac{q^{n-1}(\mathbf{f})}{p(\mathbf{f}|\theta_{n-1})}.
\end{align}
We  then approximate the KL divergence as:
\begin{align} \label{KL divergence}
    &\text{KL}[q^n(\mathbf{f}) || p(\mathbf{f}| \boldsymbol{\psi}_{1:n})] 
    = \int q^n(\mathbf{f}) \log \frac{q^n(\mathbf{f})}{p(\mathbf{f}| \boldsymbol{\psi}_{1:n})}d\mathbf{f}  \\
    &\approx\int q^n(\mathbf{f}) \log \frac{q^{n}(\mathbf{f})}{\frac{Z_{2}(\theta_{n-1})}{Z_{1}(\theta_{n})}p(\mathbf{f}|\theta_{n})p(\boldsymbol{\psi}_{n}| \mathbf{f})\frac{q^{n-1}(\mathbf{f})}{p(\mathbf{f}|\theta_{n-1})}}d\mathbf{f}\nonumber\\
    &=\log \frac{Z_{1}(\theta_{n})}{Z_{2}(\theta_{n-1})} + \int q^n(\mathbf{f}) \log \frac{q^{n}(\mathbf{u}_{n})p(\mathbf{u}_{n-1}|\theta_{n-1})}{p(\mathbf{u}_{n}|\theta_n)p(\boldsymbol{\psi}_{n}| \mathbf{f})q^{n-1}(\mathbf{u}_{n-1})}d\mathbf{f}.\nonumber
\end{align}
Here, the first term can be used to approximate the online log-marginal likelihood (as $Z_{1}(\theta_{n})/Z_{2}(\theta_{n-1}) \approx p(\boldsymbol{\psi}_{n}|\boldsymbol{\psi}_{1:n-1})$), but it is analytically intractable. 
The second term, by the non-negativity property of KL divergence, serves as a negative lower bound of the online log-marginal likelihood, known as the online ELBO.
After tedious algebraic manipulation (see Appendix \ref{Online ELBO Derivation} for detailed derivations), we can reorganize the online ELBO as follows:
\begin{align}\label{online ELBO}
\mathcal{L}_{\text{online}}=& \text{KL}[q^{n}(\mathbf{u}_{n})||p(\mathbf{u}_{n}|\theta_n)]-\sum_{\psi_i\in \boldsymbol{\psi}_n}\mathbb{E}_{q^n_{\mathbf{u}}(f_i)}[\log p(\psi_i|f_i)] \nonumber\\
&+ \text{KL}[q^{n}(\mathbf{u}_{n-1})||q^{n-1}(\mathbf{u}_{n-1})] \nonumber\\
&- \text{KL}[q^{n}(\mathbf{u}_{n-1})||p(\mathbf{u}_{n-1}|\theta_{n-1})].
\end{align}
The first two terms correspond to the standard variational lower bound for the current batch $\boldsymbol{\psi}_n$ analogous to \eqref{SVGP ELBO}, while the last two terms incorporate information from the previous posterior approximation $q^{n-1}(\mathbf{u}_{n-1})$. This decomposition thus explicitly separates contributions from past knowledge and the incremental update mechanism.

\subsubsection{Hybrid Objective Function with Memory Subset}\label{Hybrid Objective}
The effectiveness of the bound in \eqref{online ELBO} relies on accurate $q^{n-1}(\mathbf{u}_{n-1})$, but this accuracy can gradually deteriorate due to error accumulation from approximate online updates.
To address this issue and mitigate catastrophic forgetting, we incorporate a representative memory subset of previous measurements $\mathcal{M}=\{\mathbf{X}_{\mathcal{M}},\boldsymbol{\psi}_{\mathcal{M}}\} \subset \mathcal{D}_{1:n-1}$ (with size $N_\mathcal{M} = |\mathcal{M}|$) into the online variational bound calculation.
Specifically,  we form a ``pseudo-batch''  from the union of the new measurements and the memory $\mathcal{D}_n \cup \mathcal{M}$, and  construct for it an auxiliary ELBO analogous to the standard ELBO in \eqref{SVGP ELBO}:
\begin{align}\label{Pseudo-Batch ELBO}
    \mathcal{L}&_{\text{pseudo-batch}} \\
    =& \text{KL}[q^{n}(\mathbf{u}_{n})||p(\mathbf{u}_{n}|\theta_n)]-\sum_{\psi_i\in {\boldsymbol{\psi}_n} \cup \boldsymbol{\psi}_\mathcal{M}}\mathbb{E}_{q^n_{\mathbf{u}}(f_i)}[\log p(\psi_i|f_i)]. \nonumber
\end{align}
Then, we propose a hybrid objective that integrates the online ELBO \eqref{online ELBO} with the auxiliary pseudo-batch ELBO \eqref{Pseudo-Batch ELBO}. 
By introducing weighting coefficients $\mu_1$ and $\mu_2$ to balance the influence of the previous posterior and the memory, respectively, the new hybrid objective is given by:
\begin{align}\label{hybrid objective}
    &\mathcal{L}_{\text{hybrid}}=\text{KL}[q^{n}(\mathbf{u}_{n})||p(\mathbf{u}_{n}|\theta_n)]-\nonumber\\ 
    &\sum_{\psi_i\in \boldsymbol{\psi}_n}\mathbb{E}_{q^n_{\mathbf{u}}(f_i)}[\log p(\psi_i|f_i)] +\mu_1\big(\text{KL}[q^{n}(\mathbf{u}_{n-1})||q^{n-1}(\mathbf{u}_{n-1})]\nonumber\\&- \text{KL}[q^{n}(\mathbf{u}_{n-1})||p(\mathbf{u}_{n-1}|\theta_{n-1})]\big) \nonumber\\&-\mu_2\sum_{\psi_i\in \boldsymbol{\psi}_\mathcal{M}}\mathbb{E}_{q^n_{\mathbf{u}}(f_i)}[\log p(\psi_i|f_i)].
\end{align}

\subsubsection{Variational Distribution Optimization}
We minimize the hybrid objective with respect to the variational distribution $q^n(\mathbf{u}_n)$, the inducing locations $\mathbf{Z}_n$, and the hyperparameters $\theta_n$.
First,  for fixed inducing locations and hyperparameters, by taking the derivative of the hybrid objective $\mathcal{L}_{\text{hybrid}}$ with respect to $q^n(\mathbf{u}_n)$ and enforcing normalization via a Lagrange multiplier $\lambda$, i.e., $\frac{d\mathcal{L}_{\text{hybrid}}}{dq^n(\mathbf{u}_n)} + \lambda=0$, we derive the optimal variational distribution $q_{\text{opt}}^n(\mathbf{u}_n)$ as (see Appendix \ref{optimal form of q^n(u_n)} for detailed derivations):
\begin{align}\label{optimal q(u)}
    &q_{\text{opt}}^n(\mathbf{u}_n)\\
    \propto& p(\mathbf{u}_n)\exp\Big( \int p(\mathbf{f}|\mathbf{u}_n)\log p(\boldsymbol{\psi}_n|\mathbf{f}) d\mathbf{f} \nonumber\\ &+\mu_1\int p(\mathbf{u}_{n-1}|\mathbf{u}_n)\log\frac{q^{n-1}(\mathbf{u}_{n-1})}{p(\mathbf{u}_{n-1}|\theta_{n-1})}d\mathbf{u}_{n-1} \nonumber\\
    &+ \mu_2\int p(\mathbf{f}_m|\mathbf{u}_n)\log p(\boldsymbol{\psi}_{\mathcal{M}}|\mathbf{f}_m) d\mathbf{f}_m \Big) \nonumber\\
    \propto & p(\mathbf{u}_n)\mathcal{N}(\hat{\boldsymbol{\psi}}; \mathbf{K}_{\mathbf{\hat{f}}\mathbf{u}_n}\mathbf{K}_{\mathbf{u}_n\mathbf{u}_n}^{-1}\mathbf{u}_n, \Sigma_{\hat{\boldsymbol{\psi}}}) \nonumber\\
    =& \mathcal{N}\big(\mathbf{u}_n|\mathbf{K}_{\mathbf{u}_n\mathbf{\hat{f}}}(\mathbf{K}_{\mathbf{\hat{f}}\mathbf{u}_n}\mathbf{K}_{\mathbf{u}_n\mathbf{u}_n}^{-1}\mathbf{K}_{\mathbf{u}_n\mathbf{\hat{f}}}+\Sigma_{\hat{\boldsymbol{\psi}}})^{-1}\hat{\boldsymbol{\psi}}, \nonumber\\
    &\mathbf{K}_{\mathbf{u}_n\mathbf{u}_n}-\mathbf{K}_{\mathbf{u}_n\mathbf{\hat{f}}}(\mathbf{K}_{\mathbf{\hat{f}}\mathbf{u}_n}\mathbf{K}_{\mathbf{u}_n\mathbf{u}_n}^{-1}\mathbf{K}_{\mathbf{u}_n\mathbf{\hat{f}}}+\Sigma_{\hat{\boldsymbol{\psi}}})^{-1}\mathbf{K}_{\mathbf{\hat{f}}\mathbf{u}_n}\big)\label{q opt(u)}
\end{align}
where 
\begin{align}
    \hat{\boldsymbol{\psi}}&=\begin{bmatrix}\boldsymbol{\psi}_n \\\boldsymbol{\psi}_\mathcal{M}\\\hat{\mathbf{u}}_{n-1}\end{bmatrix},
\mathbf{K}_{\hat{\mathbf{f}}\mathbf{u}_{n}}=\begin{bmatrix}\mathbf{K_{fu_\mathnormal{n}}} \\\mathbf{K}_{\mathbf{f}_m\mathbf{u}_n}\\\mathbf{K}_{\mathbf{u}_{n-1}\mathbf{u}_{n}}\end{bmatrix}, \nonumber\\
\Sigma_{\hat{\boldsymbol{\psi}}}&=
\begin{bmatrix}
\sigma_{\psi}^{2}\mathbf{I}_{N_n} & 0 & 0 \\
0 & \frac{\sigma_{\psi}^{2}\mathbf{I}_{N_\mathcal{M}}}{\mu_2} & 0 \\
0 & 0 & \frac{\mathbf{D}_{\mathbf{u}_{n-1}}}{\mu_1}
\end{bmatrix}.
\end{align}
Here, $\mathbf{f}_m=f(\mathbf{X}_{\mathcal{M}})$, $\hat{\mathbf{u}}_{n-1}=\mathbf{D}_{\mathbf{u}_{n-1}}\mathbf{S}_{\mathbf{u}_{n-1}}^{-1}\mathbf{m}_{\mathbf{u}_{n-1}}$ and $\mathbf{D}_{\mathbf{u}_{n-1}}=(\mathbf{S}_{\mathbf{u}_{n-1}}^{-1}-\mathbf{K}_{\mathbf{u}_{n-1}\mathbf{u}_{n-1}}^{-1})^{-1}$. 

Then, we substitute this optimal  $q_{\text{opt}}^n(\mathbf{u}_n)$ back  into the hybrid objective $\mathcal{L}_{\text{hybrid}}$ in \eqref{hybrid objective}:
\begin{align}\label{L_theta}
    &\mathcal{L}_\text{hybrid} = \log\mathcal{N}(\hat{\boldsymbol{\psi}};0, \mathbf{K}_{\mathbf{\hat{f}}\mathbf{u}_n}\mathbf{K}_{\mathbf{u}_n\mathbf{u}_n}^{-1}\mathbf{K}_{\mathbf{u}_n\mathbf{\hat{f}}}+ \Sigma_{\hat{\boldsymbol{\psi}}}) + \Delta_1 + \mu_1 \Delta_2 \nonumber\\& + \frac{M}{2}(1-\mu_1)\log(2\pi) - \frac{M}{2}\log(\mu_1) + \frac{1}{2}(1-\mu_1)\log|\mathbf{D}_{\mathbf{u}_{n-1}}| \nonumber\\& +\mu_2 \Delta_3 + \frac{N_\mathcal{M}}{2}(1-\mu_2)\log(2\pi\sigma_{\psi}^2) - \frac{N_\mathcal{M}}{2}\log(\mu_2),
\end{align}
where 
\begin{align}
    &\Delta_1 = -\frac{1}{2\sigma_\psi^2}\text{tr}(\mathbf{Q_f}), \nonumber\\
    &\Delta_2 = \frac{1}{2}\big(-\log \frac{|\mathbf{S}_{\mathbf{u}_{n-1}}|}{|\mathbf{K}_{\mathbf{u}_{n-1}\mathbf{u}_{n-1}}^{ -1}||\mathbf{D}_{\mathbf{u}_{n-1}}|} - \text{tr}(\mathbf{D}_{\mathbf{u}_{n-1}}^{-1}\mathbf{Q}_{\mathbf{u}_{n-1}}) \nonumber \\&+\mathbf{m}_{\mathbf{u}_{n-1}}^\top (\mathbf{S}_{\mathbf{u}_{n-1}}^{-1} \mathbf{D}_{\mathbf{u}_{n-1}} \mathbf{S}_{\mathbf{u}_{n-1}}^{-1} - \mathbf{S}_{\mathbf{u}_{n-1}}^{-1})\mathbf{m}_{\mathbf{u}_{n-1}} \nonumber+M\log(2\pi)\big), \nonumber\\
    &\Delta_3 = -\frac{1}{2\sigma_\psi^2}\text{tr}(\mathbf{Q}_{\mathbf{f}_m}), \nonumber
\end{align}
with $\mathbf{Q}_{\mathbf{f}} = \mathbf{K_{ff}} - \mathbf{K}_{\mathbf{f}\mathbf{u}_n} \mathbf{K}_{\mathbf{u}_n\mathbf{u}_n}^{-1} \mathbf{K}_{\mathbf{u}_n\mathbf{f}}$, and  $\text{tr}(\cdot)$ denotes the matrix trace operator.

Next, we optimize the inducing point locations $\mathbf{Z}_n$ and the hyperparameters $\theta_n$.
To initialize $\mathbf{Z}_n$ at step $n$, we retain a specified proportion of inducing points from the previous set $\mathbf{Z}_{n-1}$ and randomly select the remaining proportion from the new measurement locations in $\mathcal{D}_n$. This initialization strategy allows the model to retain informative points from past data while incorporating information from new measurements. 
Following this initialization, we further refine the inducing locations $\mathbf{Z}_n$ and the hyperparameters $\theta_n$ by maximizing \eqref{L_theta} with gradient descent.



Finally, the variational posterior $q^n(\mathbf{f})$ provides an updated approximation of the exact posterior $p(\mathbf{f}| \boldsymbol{\psi}_{1:n})$, and the latent function estimate $\hat{f}^{n-1}$ is updated to $\hat{f}^{n}$.
With the optimal variational parameters $\mathbf{m}_{\mathbf{u}_n}$, $\mathbf{S}_{\mathbf{u}_n}$ associated with $q^n(\mathbf{f})$, we can estimate the mean $\boldsymbol{\mu}_n^*$ and variance $\mathbf{\Sigma}_n^*$ of RSS values $\boldsymbol{\psi}_n^*$ at non-measured points $\mathbf{X}_n^*$ as follows:
\begin{align}
&\boldsymbol{\mu}_n^*=\mathbf{K}_{\boldsymbol{\psi}_n^*{\mathbf{u}_n}}\mathbf{K}_{\mathbf{u}_n\mathbf{u}_n}^{-1}\mathbf{m}_{\mathbf{u}_n},\label{eq:prediction_3}\\
&\mathbf{\Sigma}_n^*=\mathbf{K}_{\boldsymbol{\psi}_n^*\boldsymbol{\psi}_n^*}-\mathbf{K}_{\boldsymbol{\psi}_n^*\mathbf{u}_n}\mathbf{K}_{\mathbf{u}_n\mathbf{u}_n}^{-1}\left(\mathbf{K}_{\mathbf{u}_n\mathbf{u}_n}-\mathbf{S}_{\mathbf{u}_n}\right)\mathbf{K}_{\mathbf{u}_n\mathbf{u}_n}^{-1}\mathbf{K}_{\mathbf{u}_n\boldsymbol{\psi}_n^*}.\label{eq:prediction_4}
\end{align}
The per-update optimization complexity of our method is $\mathcal{O}((N_n+N_\mathcal{M})M^2)$ and does not grow with the number of processed batches.

Regarding practical implementation, both the optimization of the objective function $\mathcal{L}_{\text{hybrid}}$  in \eqref{L_theta} and the evaluation of the  mean $\boldsymbol{\mu}_n^*$ and variance $\mathbf{\Sigma}_n^*$ in \eqref{eq:prediction_3} and \eqref{eq:prediction_4} involve matrix inversions and log-determinants that can be numerically unstable. To address these potential issues, we derive and present numerically stable computational forms for both the hybrid objective and the predictive distribution in Appendix~\ref{L calculate} and Appendix~\ref{Prediction}, respectively.

\subsection{Grid-assisted Online Inducing Points Selection}\label{GOIPS}
In M-OSVGP, inducing points are initialized through random sampling, and their locations are subsequently refined via gradient-based optimization. This initialization strategy has two potential drawbacks.
First, it does not take into account the spatial properties of new measurements. If new measurements arrive far from the existing inducing set, the gradient-based optimization may converge slowly and even be trapped in a less favorable local solution. Second, the random selection process may sample redundant or closely spaced points, which would reduce the efficiency of the limited inducing point set.

To address these issues, we propose the GOIPS algorithm, which replaces the random initialization in M-OSVGP with a spatially aware, grid-assisted strategy. For each new batch, GOIPS adaptively refines the inducing set by adding points that cover previously unrepresented regions and removing redundant points to maintain a bounded size. The algorithm selects inducing points that are close to the measurements in the reproducing kernel Hilbert space (RKHS) \cite{GalyFajou2021} to ensure accurate approximation, while maintaining sufficient diversity to avoid redundancy. This results in a geographically representative set that preserves both coverage and representativeness. The approach provides a controlled trade-off between computational cost and model accuracy and establishes a stronger starting point for subsequent gradient-based refinement.

\begin{algorithm}
\caption{Grid-assisted Online Inducing Point Selection}
\begin{algorithmic}[1]
\State \textbf{Input:} New  measurement locations $\mathbf{X}_n=[\boldsymbol{x}_{n, i}]_{i=1}^N$, current inducing points $\mathbf{Z}_{n-1} = [\boldsymbol{z}_{n-1,i}]_{i=1}^M$, kernel function $k(\cdot,\cdot)$, threshold $\rho$, maximum and minimum number of inducing points $M_{max}$, $M_{min}$.
\State Initialize $\mathbf{Z}_n \leftarrow \mathbf{Z}_{n-1}$.
\State Define $g(\boldsymbol{x}_{n, i})$ as the function mapping a point to its grid cell index.
\State Define $\mathcal{G}_{\mathbf{Z}_n} = {\{g(\boldsymbol{z}_{n, i})|\boldsymbol{z}_{n,i} \in \mathbf{Z}_n\}}$ as the set of occupied grid cells.
\For{each $\boldsymbol{x}_{n,i}\in \mathbf{X}_n$}
\State $g_{\boldsymbol{x}_{n,i}} \gets g(\boldsymbol{x}_{n,i})$
\If{$g_{\boldsymbol{x}_{n,i}} \notin \mathcal{G}_{\mathbf{Z}_n}$}
    \State $\mathbf{Z}_n \gets \mathbf{Z}_n \cup \{\boldsymbol{x}_{n,i}\}$
    \State $\mathcal{G}_{\mathbf{Z}_n} \gets \mathcal{G}_{\mathbf{Z}_n} \cup \{g_{\boldsymbol{x}_{n,i}}\}$
\Else
    \State $M_{g_{\boldsymbol{x}_{n,i}}} \gets$ number of inducing points in grid $g_{\boldsymbol{x}_{n,i}}$
    \State $\rho_g \gets \rho / M_{g_{\boldsymbol{x}_{n,i}}}$
    \State $d \gets \max_{\boldsymbol{z}_{n,i} \in \mathbf{Z}_n} k(\boldsymbol{x}_{n,i}, \boldsymbol{z}_{n,i})$
    \If{$d < \rho_g$}
        \State $\mathbf{Z}_n \gets \mathbf{Z}_n \cup \{\boldsymbol{x}_{n,i}\}$
    \EndIf
\EndIf
\EndFor
\If{$|\mathbf{Z}_n| > M_{max}$}
\While{$|\mathbf{Z}_n| > M_{min}$}
    \State Compute correlation matrix $\mathbf{K}_{\mathbf{Z}_n\mathbf{Z}_n}$.
    \State For each $\boldsymbol{z}_{n,i} \in \mathbf{Z}_n$, compute overlap $C_i = \phantom{---}  \sum_{j \neq i} \mathbbm{1}[\mathbf{K}_{ij} > \rho]$.
    \State Define removal candidate set $\mathcal{Z}_{rem} = \{\boldsymbol{z}_{n,i} \mid C_i > 1\}$.
    \If{$\mathcal{Z}_{rem} \neq \emptyset$ }
        \State Select $\boldsymbol{z}_{rem}$ from $\mathcal{Z}_{rem}$ with probability  $P(\boldsymbol{z}_{n,i}) \propto C_i$.
    \State $\mathbf{Z}_n \gets \mathbf{Z}_n \setminus \{\boldsymbol{z}_{rem}\}$.

    \EndIf
    \EndWhile
\EndIf
\Statex  
\Return{Updated inducing point set $\mathbf{Z}_n$}
\end{algorithmic}
\label{alg:1}
\end{algorithm}

\subsubsection{Adding New Inducing Points}
To ensure good spatial coverage, the target region is partitioned into uniform grids. 
During the online learning process, for each incoming measurement location $\boldsymbol{x}_{n, i}\in \mathbf{X}_n$, the algorithm first determines its corresponding grid cell. 
If no inducing points exist in the grid, $\boldsymbol{x}_{n, i}$ is immediately added to $\mathbf{Z}_n$ to cover a new, previously unrepresented area. 
Otherwise, we assess the distinctiveness of $\boldsymbol{x}_{n, i}$ with respect to the current inducing point set in the RKHS.
We compute its maximum kernel similarity to all existing inducing points, $d=\max_{\boldsymbol{z}_{n, i} \in \mathbf{Z}_n} k(\boldsymbol{x}_{n, i}, \boldsymbol{z}_{n, i})$.
The point $\boldsymbol{x}_{n, i}$ is added to $\mathbf{Z}_n$ only if this similarity score $d$ is below an adaptive threshold $\rho_g$, which indicates that the new point provides sufficiently novel information. 
If not, the algorithm moves on to the next measurement. 
The acceptance threshold $\rho_g$ is dynamically adjusted based on the sample density of the grid: densely populated grids are assigned a lower $\rho_g$ to reduce redundancy, while sparsely populated grids have a higher $\rho_g$ to retain more representative points.
\begin{figure}[!t]
    \centering
    
    \subfloat[t][Scenario (\textit{Campus})]{\includegraphics[height=3.6cm]{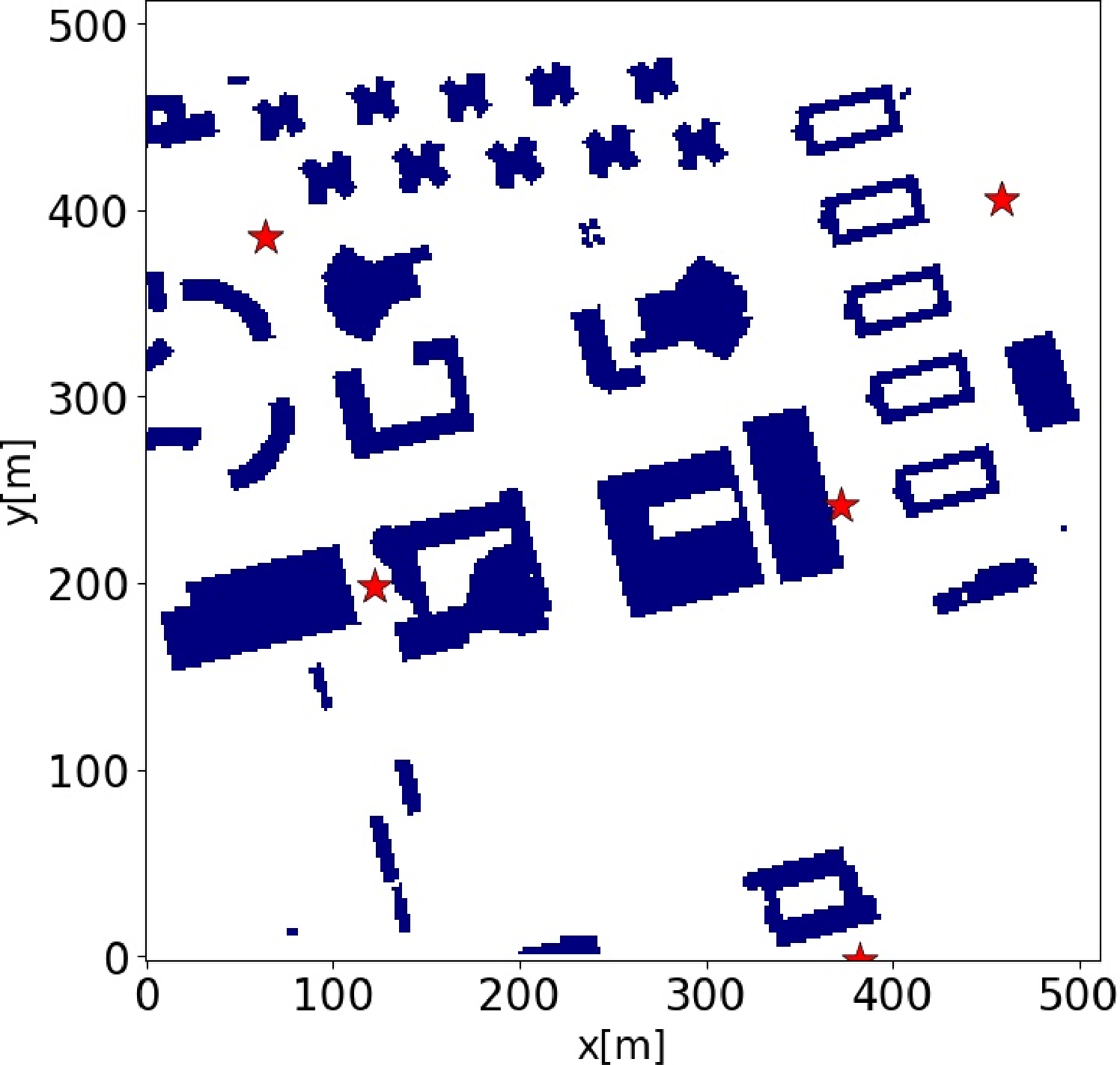}\label{fig:senario_nuaa}}
    \hfill
    \subfloat[t][True radio map (\textit{Campus})]{\includegraphics[height=3.6cm]{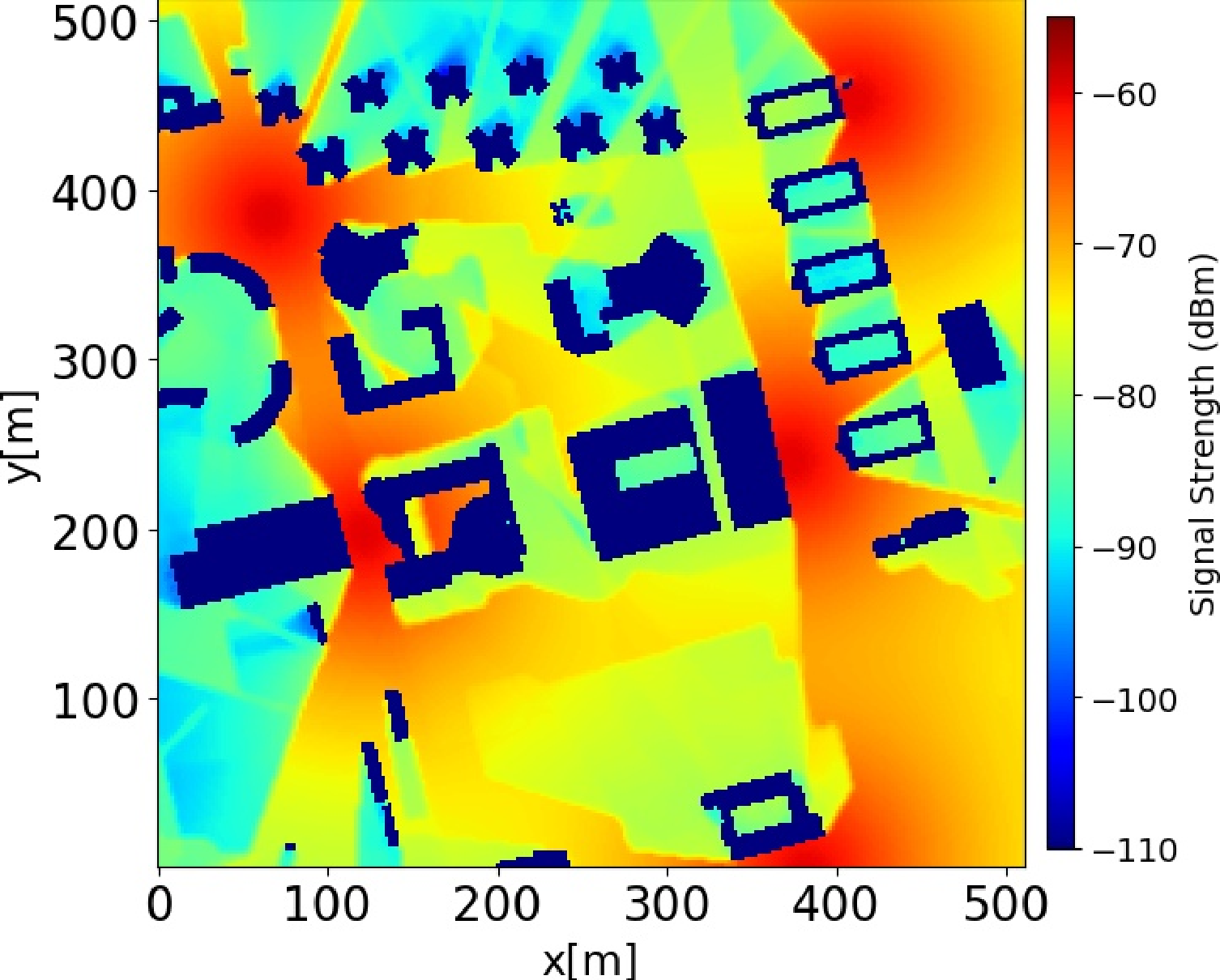}\label{fig:True_map_nuaa}}
    \hfill
    \subfloat[t][Scenario (\textit{Urban})]{\includegraphics[height=3.6cm]{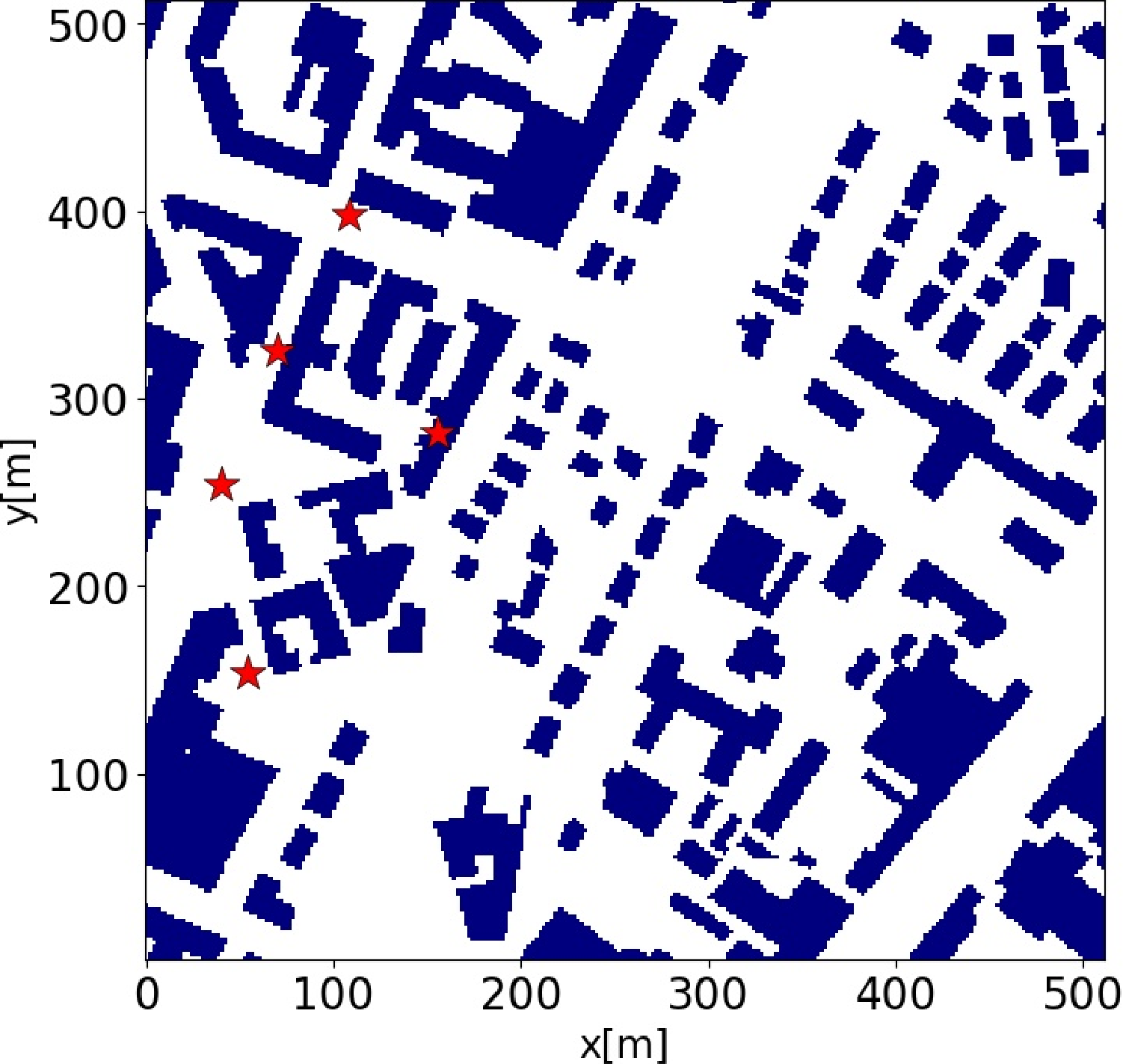}\label{fig:senario_milano}}
    \hfill
    \subfloat[t][True radio map (\textit{Urban})]{\includegraphics[height=3.6cm]{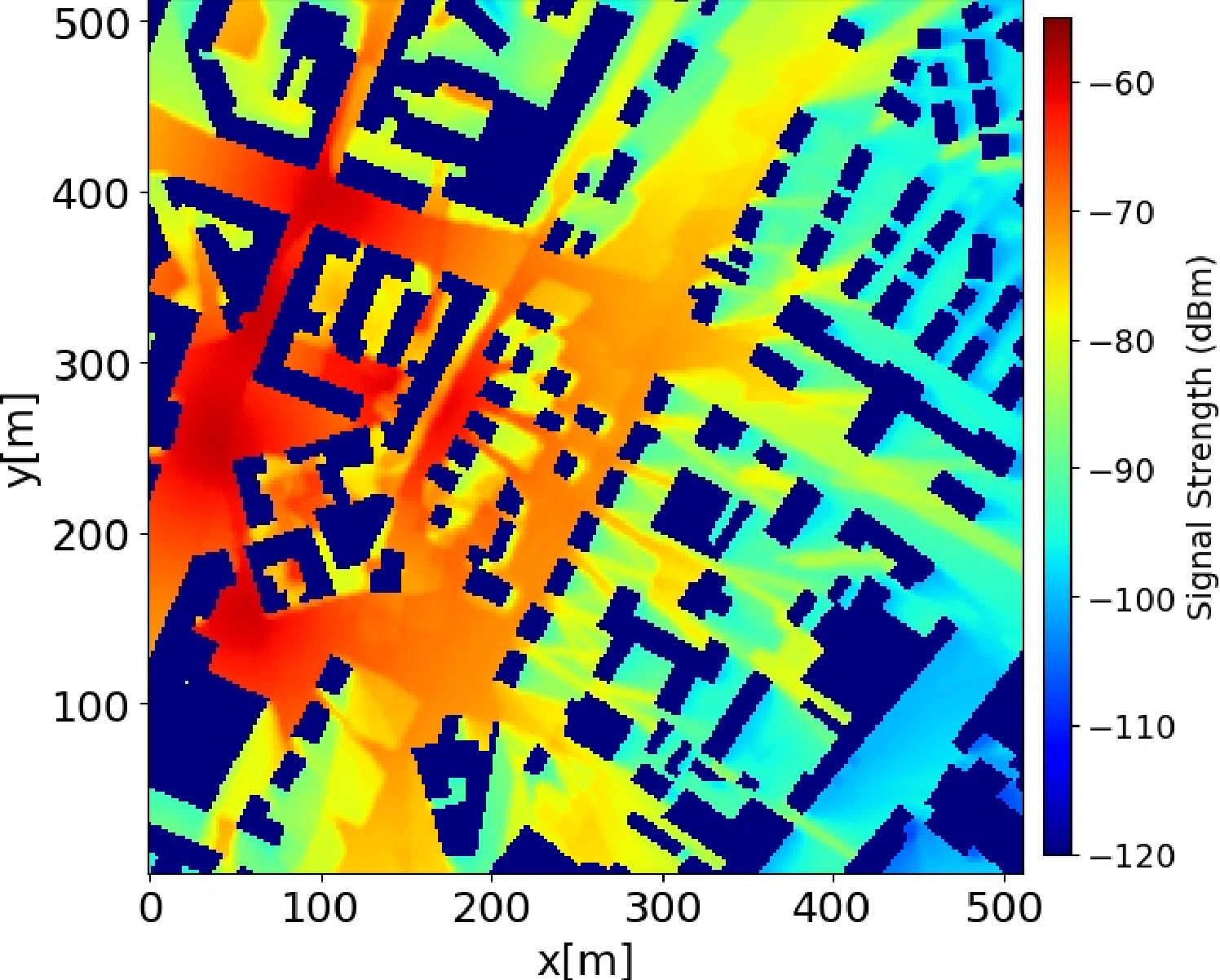}\label{fig:True_map}}

    \caption{Simulation scenario and the associated true radio map, with red stars in (a) and (c) indicating emitters' locations. }
    \label{fig:scenario}
\end{figure}

\subsubsection{Dynamic Maintenance of Inducing Points}
To maintain computational tractability, we introduce a removal strategy when the number of inducing points exceeds the predefined maximum $M_{max}$. 
This strategy identifies and removes redundant points based on their kernel correlation. 
For each inducing point $\boldsymbol{z}_{n, j}$, we first compute the ``overlap count" $C_i$, defined as the number of other inducing points with kernel similarity exceeding the threshold $\rho$. 
Points with $C_i > 1$ are considered redundant and eligible for removal. 
The removal process uses a probabilistic scheme: points with higher overlap counts have a greater probability of being removed, ensuring that the most redundant points are eliminated first. 
This process continues until no points satisfy the removal criteria, or the inducing set size is reduced to the minimum threshold $M_{min}$.
The complete procedure is described in Algorithm~\ref{alg:1}. 
It is important to note that the output of GOIPS is not the final configuration of inducing points. 
Instead, it provides an initialization that is subsequently refined through gradient descent during the variational optimization stage described in Section~\ref{Approximated Online Sparse GP}.

The proposed GOIPS is naturally well-suited for online learning from streaming measurements, as each measurement is processed only once. 
In particular, the kernel similarities $K_{x\mathbf{Z}_n}$ required for adding new inducing points are already computed during the variational GP updates, and thus incurs minimal additional cost.
The removal step constrains the number of inducing points within the range $[M_{min}, M_{max}]$, thereby keeping the computational complexity manageable.
Moreover, the use of grid-based spatial partitioning combined with density-adaptive thresholding naturally reflects the geographical properties of radio map construction, improving both the efficiency and the spatial representativeness of inducing point allocation.

\section{Simulation Results and Analysis}\label{sec:simulations}
In this section, we evaluate the performance of our proposed M-OSVGP and M-OSVGP-GOIPS methods through extensive simulations.
We first describe the experimental setup, including the simulation scenarios and evaluation metrics, and then compare our methods against existing baselines in terms of accuracy, efficiency, and uncertainty. 
We further study the effects of key hyperparameters and the effectiveness of the proposed GOIPS module.

\begin{table}[t]
\centering
\caption{Simulation parameters.}
\label{tab:Parameter}
\resizebox{0.45\textwidth}{!}{
\begin{tabular}{cc}
\toprule
\text{Simulation parameter} & \text{Value} \\
\midrule
Size of target area & $512\, \text{m} \times 512\, \text{m}$ \\
Grid size & $2\, \text{m}$ \\
Number of emitters & 5 \\
Operating frequency of emitter & $3.5\, \text{GHz}$ \\
Maximum power of an emitter & $23\, \text{dBm}$ \\
Sensors' height & $1.5\, \text{m}$ \\
Transmitting antenna height & $40\, \text{m}$ \\
\bottomrule
\end{tabular}
}
\end{table}

\subsection{Simulation Setup}
As illustrated in Fig.\ref{fig:scenario}, we consider two distinct scenarios, \textit{campus} and \textit{urban}, generated using the radio network planning software WinProp \cite{Hoppe2017} with the Dominant Path Model to generate the ground-truth RSS distribution across the map, i.e., the true radio map.
The \textit{Campus} scenario (Fig.\ref{fig:senario_nuaa} and \ref{fig:True_map_nuaa}) corresponds to the Jiangning campus of Nanjing University of Aeronautics and Astronautics, Nanjing, China, and the \textit{Urban} scenario (Fig.\ref{fig:senario_milano} and \ref{fig:True_map}) corresponds to a dense urban area in Milano, Italy.
The key simulation parameters are summarized in Table~\ref{tab:Parameter}.
For each scenario, we randomly sample the true radio map to obtain the spectrum measurements, which are divided into ten sequential batches: the first contains 600 samples ($N_1 = 600$), and each subsequent batch contains 200 samples ($N_j = 200$, $j = 2,\ldots,10$). 

\begin{figure*}[htbp]
    \centering
    \begin{subfigure}[b]{0.32\textwidth} 
        \centering
        \includegraphics[width=\textwidth]{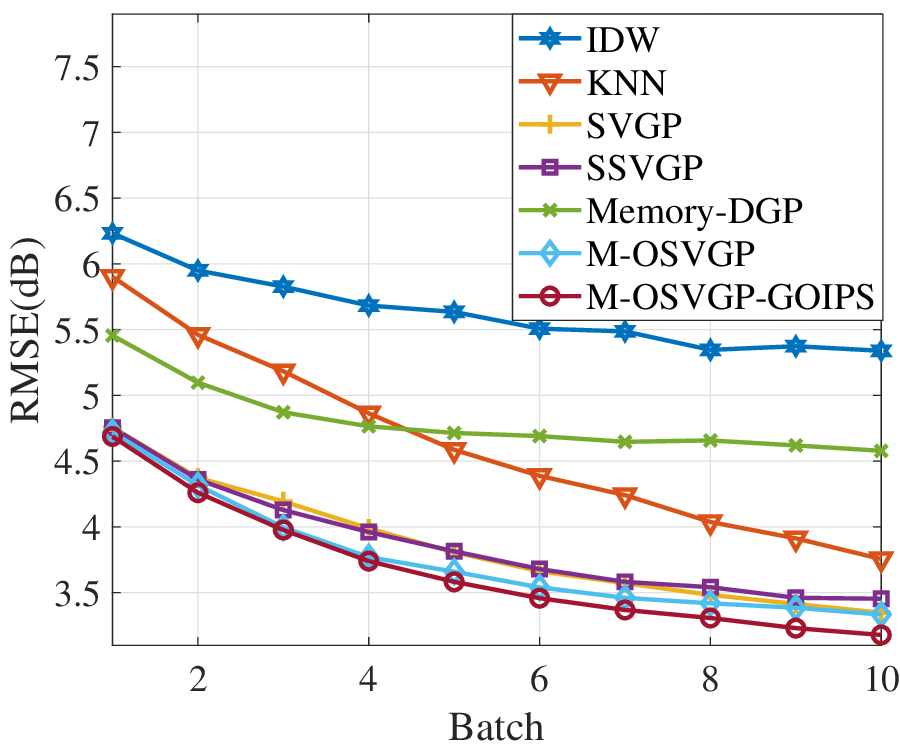}
        \caption{RMSE (\textit{Campus})}
        \label{fig:RMSE_nuaa}
    \end{subfigure}
    \hfill    
    \begin{subfigure}[b]{0.32\textwidth} 
        \centering
        \includegraphics[width=\textwidth]{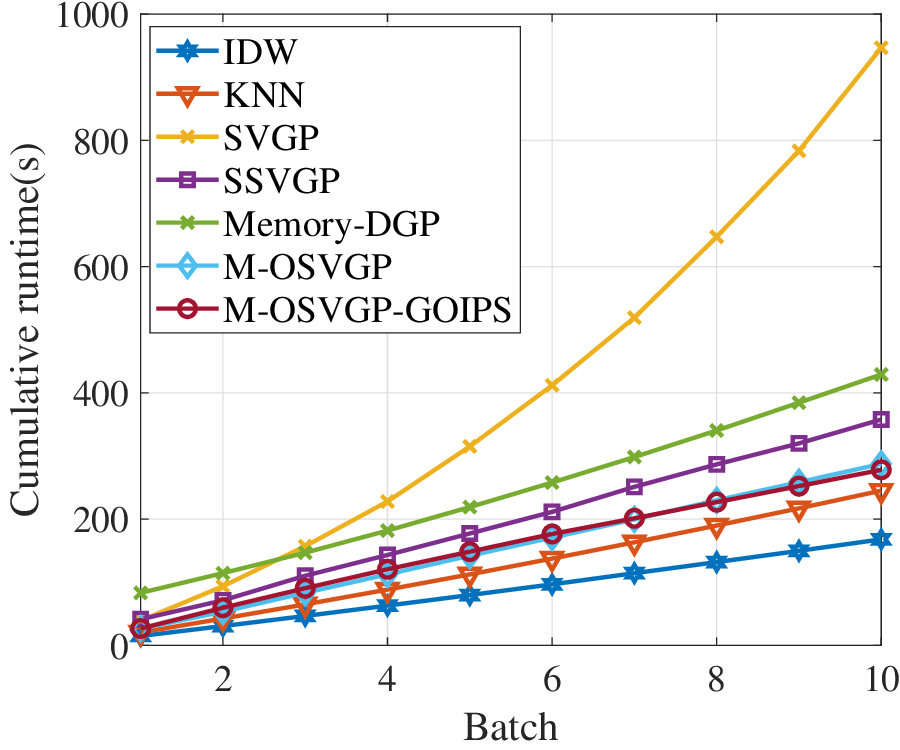}
        \caption{Cumulative runtime (\textit{Campus})}
        \label{fig:Time_nuaa}
    \end{subfigure}
    \hfill
    \begin{subfigure}[b]{0.32\textwidth} 
        \centering
        \includegraphics[width=\textwidth]{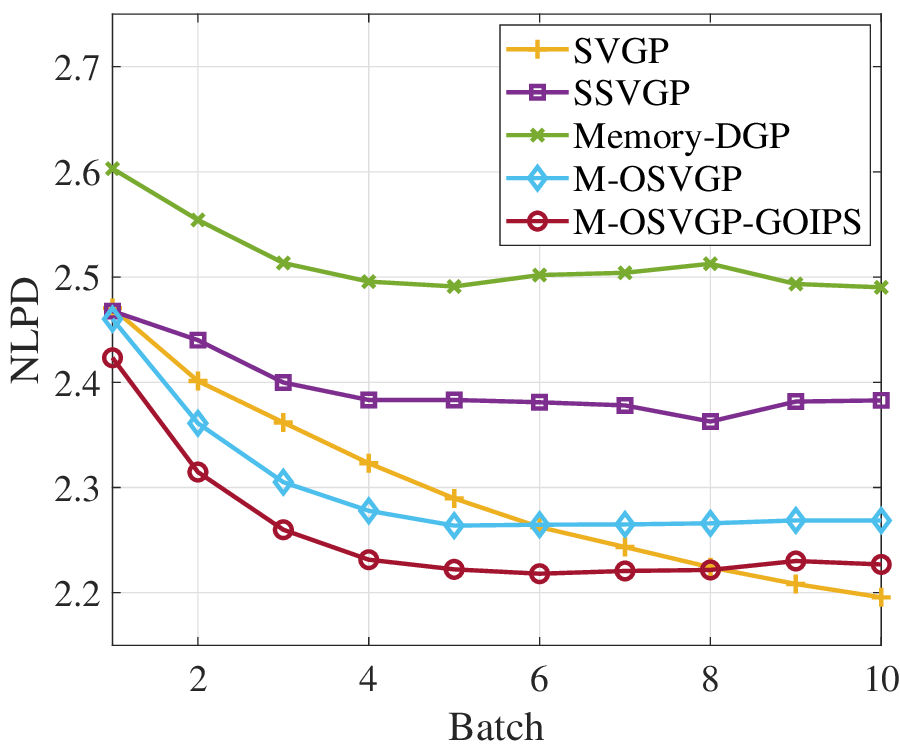}
        \caption{NLPD (\textit{Campus})}
        \label{fig:NLPD_nuaa}
    \end{subfigure}
    

    \begin{subfigure}[b]{0.32\textwidth} 
        \centering
        \includegraphics[width=\textwidth]{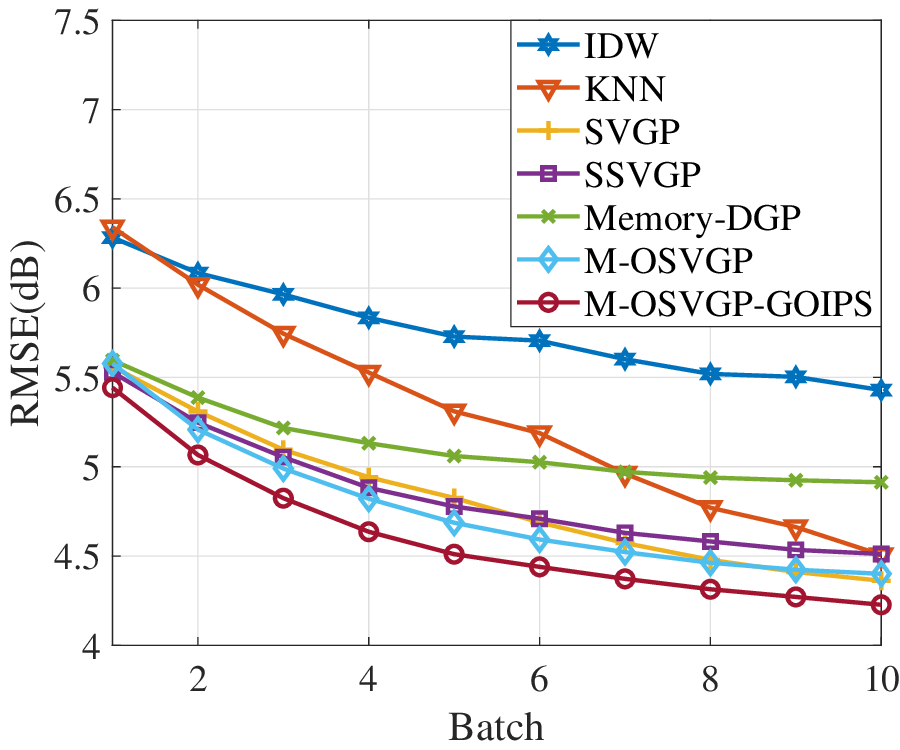}
        \caption{RMSE (\textit{Urban})}
        \label{fig:RMSE_Milano}
    \end{subfigure}
    \hfill
    \begin{subfigure}[b]{0.32\textwidth} 
        \centering
        \includegraphics[width=\textwidth]{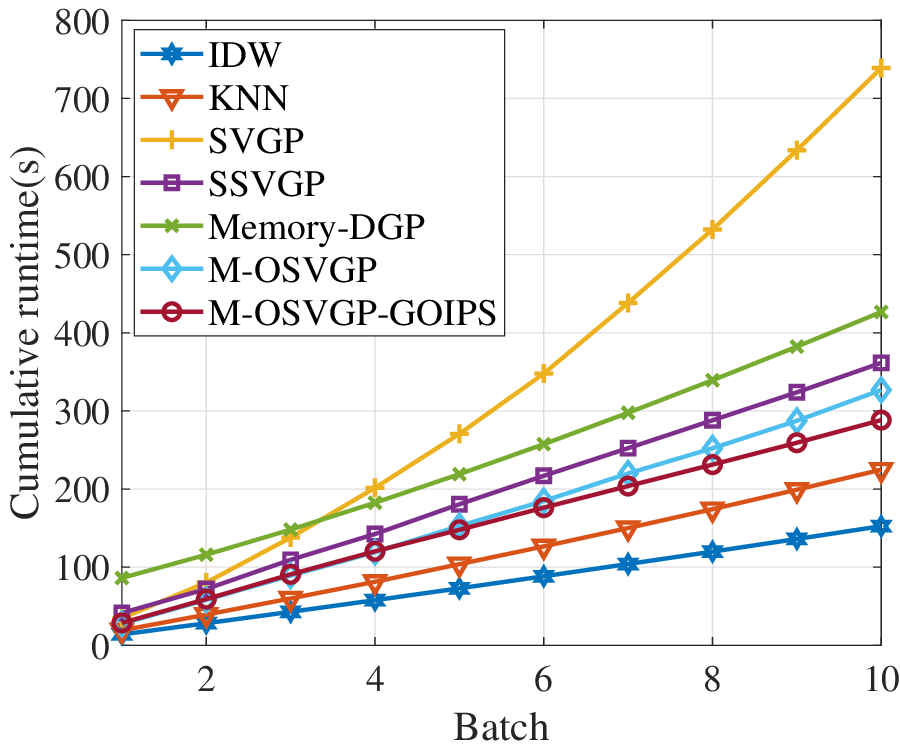}
        \caption{Cumulative runtime (\textit{Urban})}
        \label{fig:Time_Milano}
    \end{subfigure}
    \hfill
    \begin{subfigure}[b]{0.32\textwidth} 
        \centering
        \includegraphics[width=\textwidth]{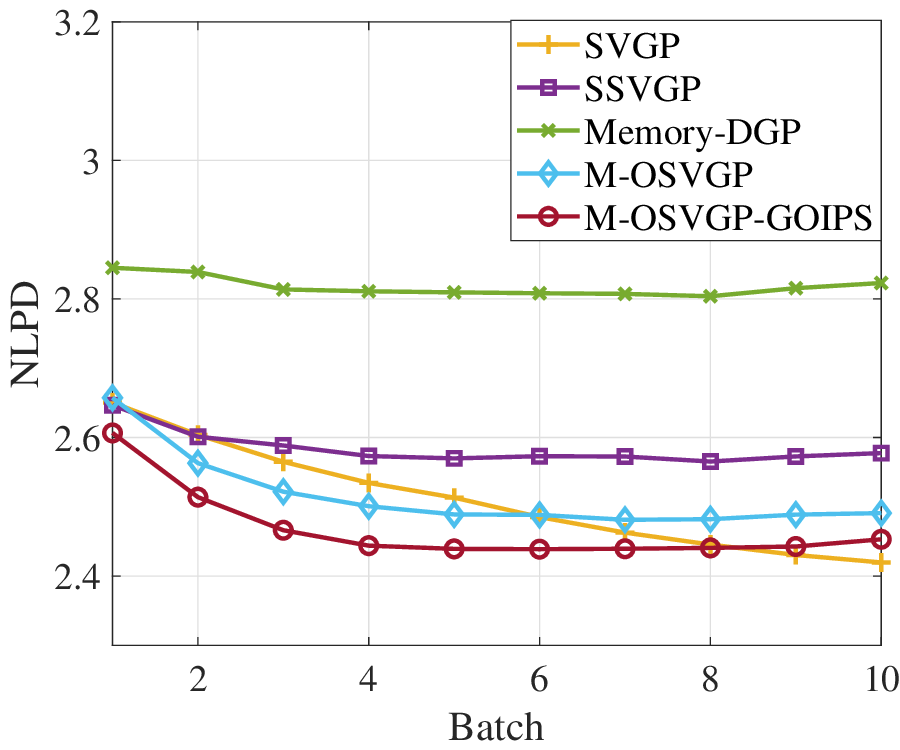}
        \caption{NLPD (\textit{Urban})}
        \label{fig:NLPD_Milano}
    \end{subfigure}
    \caption{Comparison of RMSE, cumulative runtime, and NLPD for both scenarios.}
    \label{fig:combined_metrics}
\end{figure*}

We compare the proposed M-OSVGP and M-OSVGP-GOIPS with five baseline methods: KNN, IDW, SVGP \cite{Titsias2009}, SSVGP \cite{Bui2017}, and Memory-DGP \cite{Chang2023}.
Among these, KNN, IDW, and SVGP are batch-processing methods that reconstruct the entire radio map by reprocessing all received measurements whenever a new batch arrives.
In contrast, SSVGP and Memory-DGP are online methods that incrementally update the radio map using the $n$-th batch of measurements, where Memory-DGP uses dual sparse variational GP with a memory of previous measurements.
Note that SVGP, SSVGP, and Memory-DGP are all GP-based methods, while KNN and IDW are heuristic interpolation techniques.

We evaluate the performance in terms of reconstruction accuracy, estimation uncertainty, computational complexity. 
The reconstruction accuracy is measured using the root mean square error (RMSE), defined as $\text{RMSE}=\sqrt{\frac{1}{N}\sum_{i=1}^{N}||\psi(\boldsymbol{x}_{i})-\tilde{\psi}(\boldsymbol{x}_{i})||_2^2}$, where $\psi(x_{i})$ and $\tilde{\psi}(\boldsymbol{x}_{i})$ denote the true and estimated RSS at location $\boldsymbol{x}_{i}$, respectively, and $N$ is the total number of non-measured points.
The estimate uncertainty is measured using the negative log predictive density (NLPD) \cite{Williams2006}, defined as $\text{NLPD} = \frac{1}{N} \sum_{i=1}^{N} -\log p(\psi(\boldsymbol{x}_{i})|\tilde{\psi}(\boldsymbol{x}_{i}))$, where lower NLPD indicates better uncertainty calibration and more reliable confidence estimates.
The computational complexity is measured using the cumulative runtime $T_{n}$, defined as the total runtime required for training the model and estimating the RSS values at non-measured points from the first batch to the $n$-th batch. 

All GP-based methods employ an additive kernel comprising three Mat\'{e}rn kernels with different smoothness parameters: Mat\'{e}rn 1/2, Mat\'{e}rn 3/2, and Mat\'{e}rn 5/2 \cite{Williams2006}.
Sparse variational methods use $M=300$ inducing points.
For the M-OSVGP method, the hybrid objective function weights are $\mu_1=\mu_2=1$, and the size of the memory subset is $N_\mathcal{M}=500$, randomly sampled from previously received measurements. Its inducing point initialization strategy retains 70\% of the points from the previous step and randomly selects the remaining 30\% from new measurements.
The M-OSVGP-GOIPS method employs the GOIPS algorithm with a grid resolution of $25 \times 25$ meters, acceptance threshold $\rho = 0.9$, and the maximum and minimum number of inducing points of $M_{max} = 350$ and $M_{min} = 250$, respectively.
Simulations are conducted on an Intel Xeon Platinum 8352V CPU with 32GB RAM and results are averaged over 30 independent runs.

\subsection{Performance Comparison}
In Fig.~\ref{fig:combined_metrics}, we compare the RMSE, the cumulative runtime, and the NLPD for different methods in both scenarios. 
Overall, our proposed methods achieve the highest reconstruction accuracy among all baselines while providing reliable uncertainty quantification and maintaining efficient computation compared to other GP-based approaches.

Specifically, regarding reconstruction accuracy, as shown in Figs.~\ref{fig:RMSE_nuaa} and \ref{fig:RMSE_Milano}, our proposed M-OSVGP and M-OSVGP-GOIPS consistently achieve the lowest RMSE in both scenarios.
In the \textit{campus} scenario, the baseline M-OSVGP reduces the RMSE by up to 4.9\%, 4.3\%, and 25.6\% compared to SVGP, SSVGP, and Memory-DGP, respectively; in the \textit{urban} scenario, the corresponding reductions are up to 3.1\%, 2.7\%, and 11.2\%.
The GOIPS algorithm further improves performance, with M-OSVGP-GOIPS achieving  up to 1.8\% additional RMSE reduction compared with M-OSVGP.

For computational efficiency, as shown in Figs.~\ref{fig:Time_nuaa} and \ref{fig:Time_Milano}, our proposed online methods are substantially more efficient than other GP-based models. Specifically, in the \textit{campus} scenario, M-OSVGP reduces the runtime by up to 71.3\%, 36.1\%, and 68.3\% compared to SVGP, SSVGP, and Memory-DGP, respectively, and similar trends are observed in the \textit{urban} scenario, with reductions of up to 58.2\%, 33.5\%, and 68.1\%. 
Notably, M-OSVGP-GOIPS not only improves accuracy but also further reduces computational time compared to M-OSVGP. 
Although the heuristic methods, KNN and IDW are faster, they achieve lower reconstruction accuracy and cannot provide principled uncertainty estimates  (and are thus excluded from the following uncertainty analysis).

For estimate uncertainty, as seen in Figs.~\ref{fig:NLPD_nuaa} and \ref{fig:NLPD_Milano}, our proposed methods yield lower NLPD values than the other online baselines, SSVGP and Memory-DGP. 
While the batch SVGP method shows a steady decrease in NLPD due to its use of all accumulated measurements, our proposed methods approach the uncertainty calibration of batch processing through our hybrid objective with memory subset, while retaining online efficiency.
The M-OSVGP-GOIPS method further improves these uncertainty estimates, as its spatially-aware inducing point selection better captures the underlying spatial variability. 

\begin{figure}[htbp]
    \centering
    \begin{minipage}[t]{0.147\textwidth}  
        \centering
        \includegraphics[width=\textwidth]{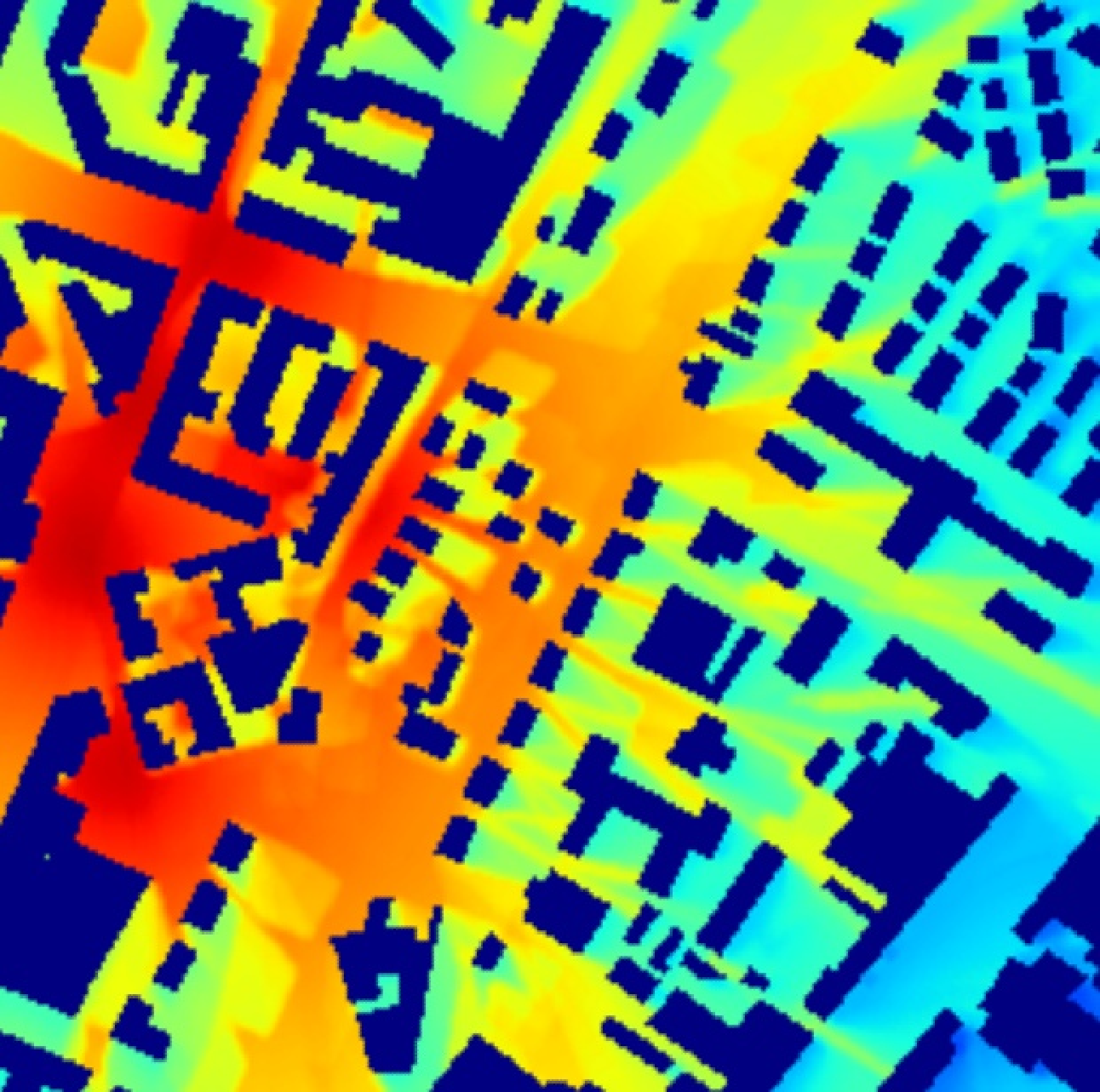} 
        \caption*{\small{True radio map}}
    \end{minipage}
    \hfill
    \begin{minipage}[t]{0.147\textwidth}  
        \centering
        \includegraphics[width=\textwidth]{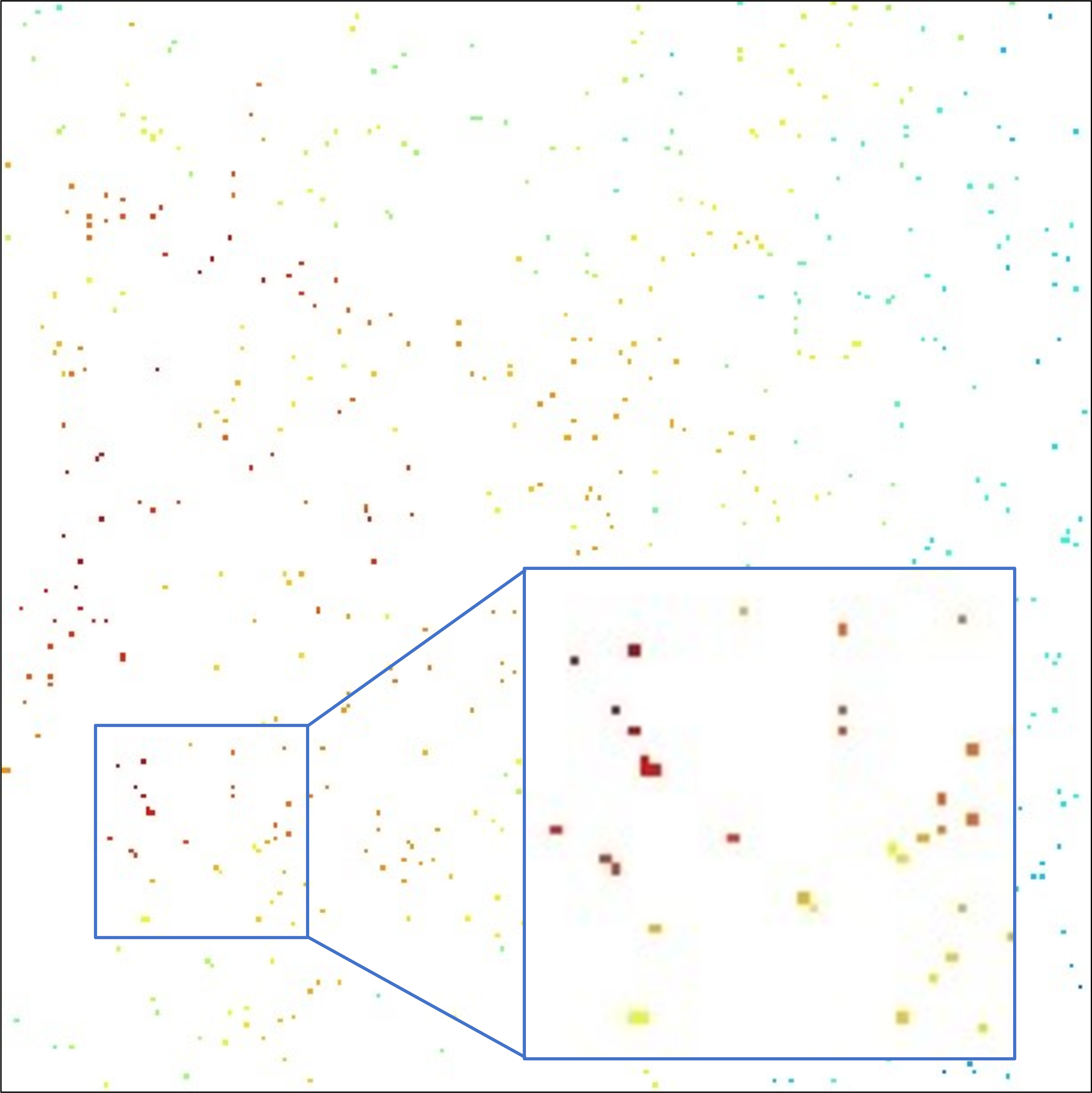} 
        \caption*{\small{Measurements}}
    \end{minipage}
    \hfill
    \begin{minipage}[t]{0.147\textwidth}  
        \centering
        \includegraphics[width=\textwidth]{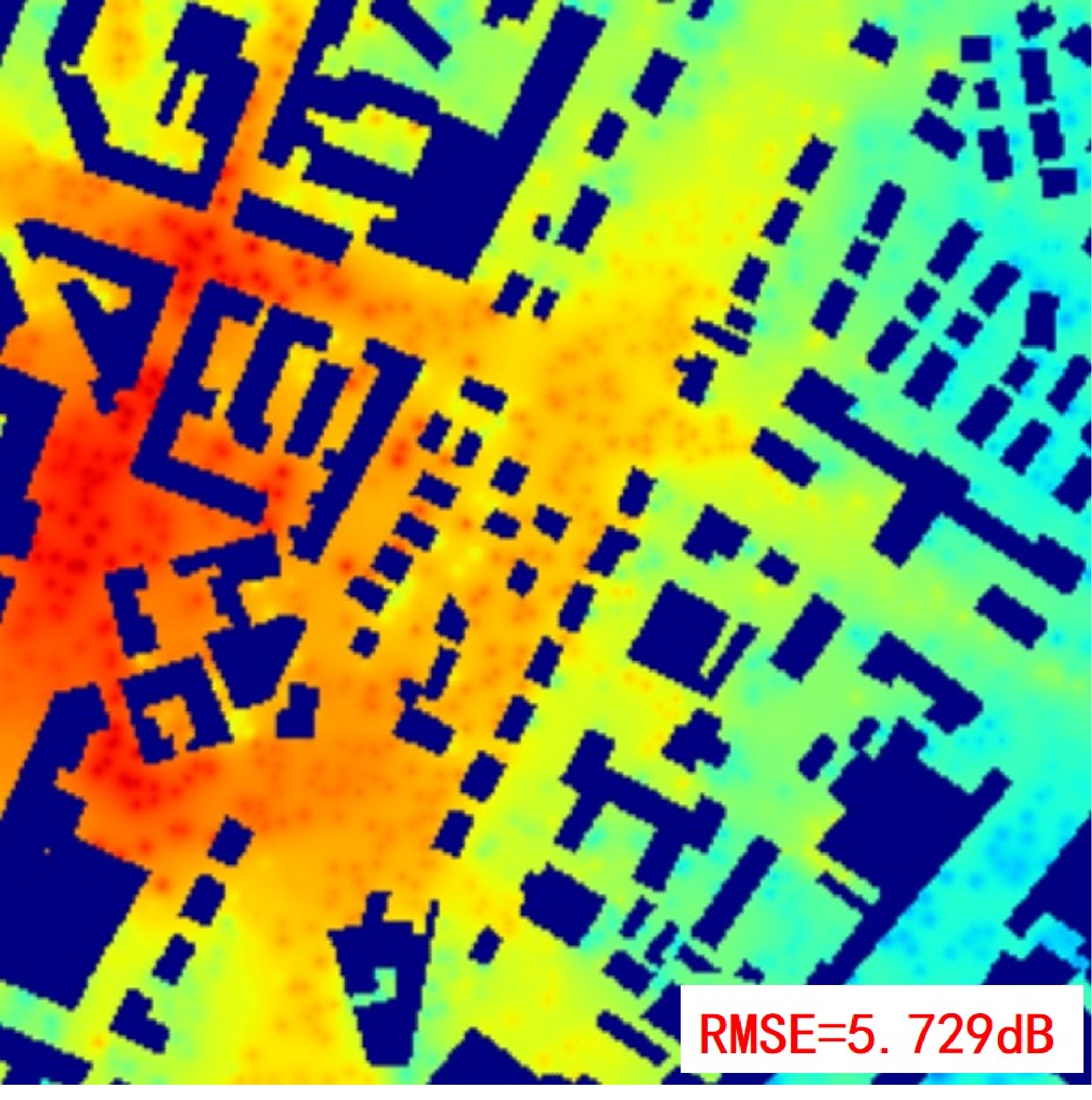}
        \caption*{\small{IDW}}
    \end{minipage}
    \hspace{-0.16cm}  
    \begin{minipage}[t]{0.027\textwidth}  
        \centering
        \includegraphics[width=\textwidth]{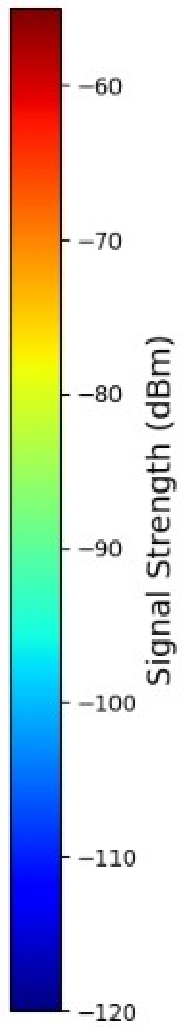} 
    \end{minipage}
    \vspace{0.1cm}
    \hfill
    \begin{minipage}[t]{0.147\textwidth}  
        \centering
        \includegraphics[width=\textwidth]{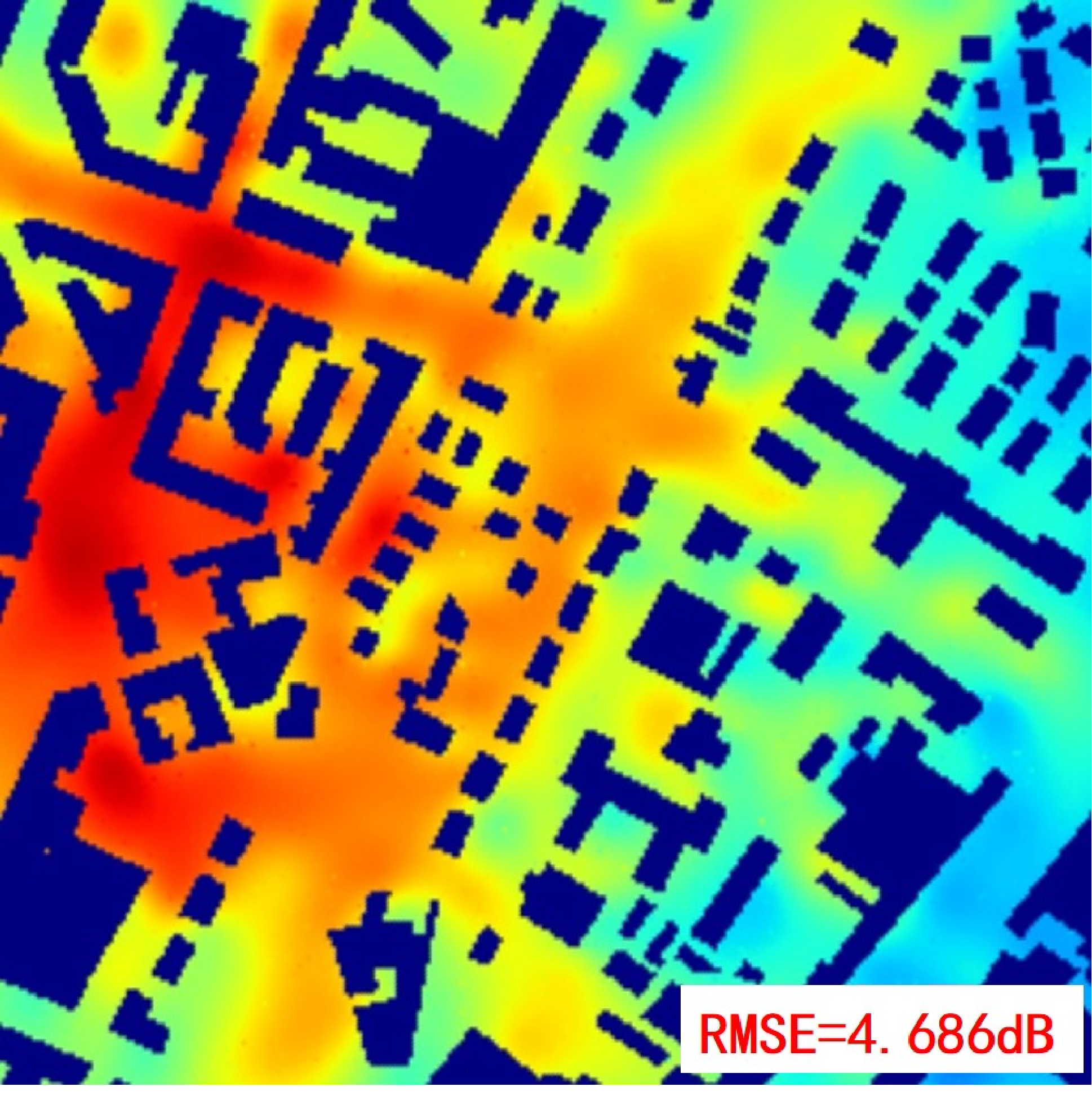} 
        \caption*{\small{M-OSVGP}}
    \end{minipage}
    \hfill
    \begin{minipage}[t]{0.147\textwidth}  
        \centering
        \includegraphics[width=\textwidth]{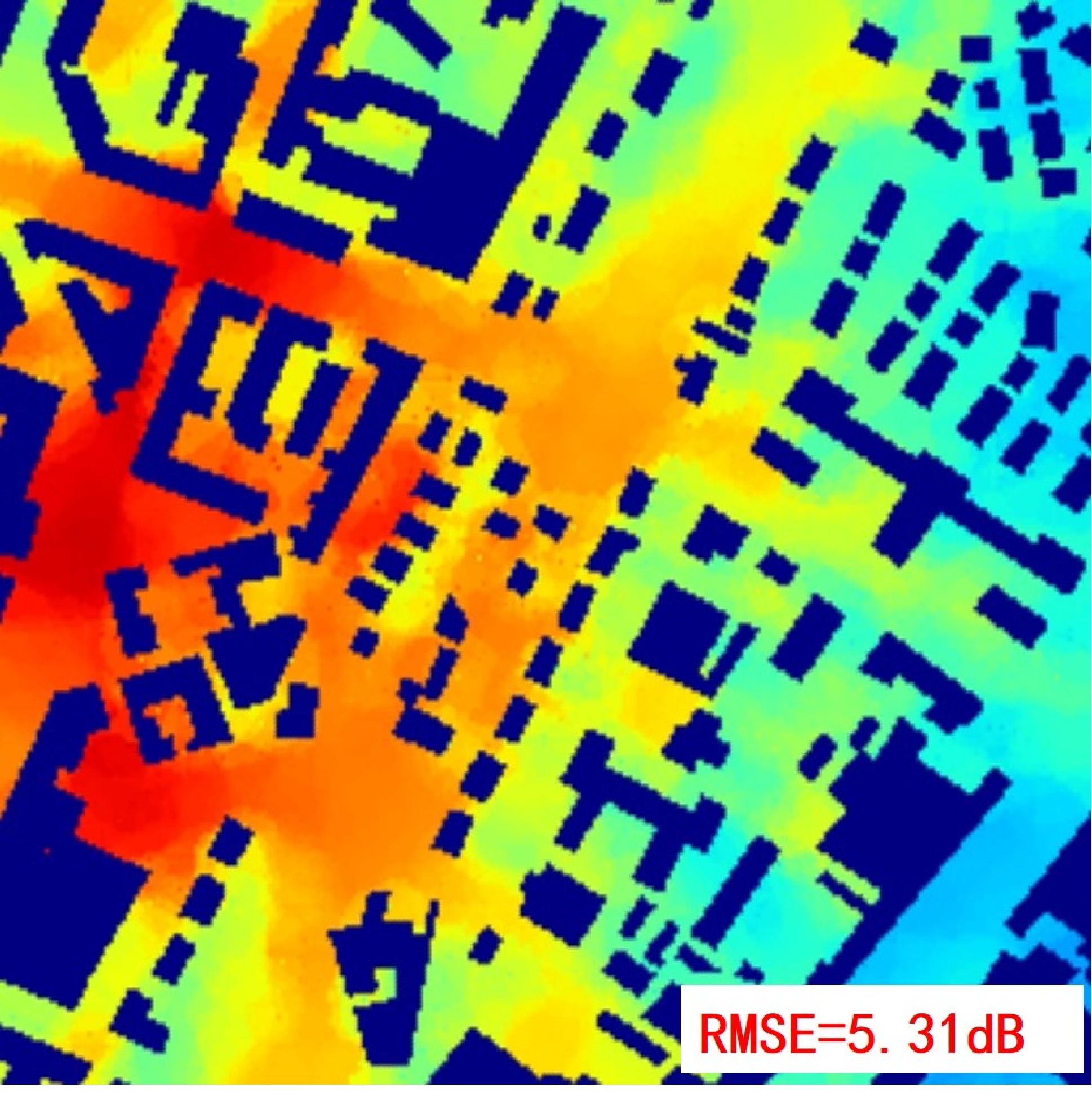} 
        \caption*{\small{KNN}}
    \end{minipage}
    \hfill
    \begin{minipage}[t]{0.147\textwidth}  
        \centering
        \includegraphics[width=\textwidth]{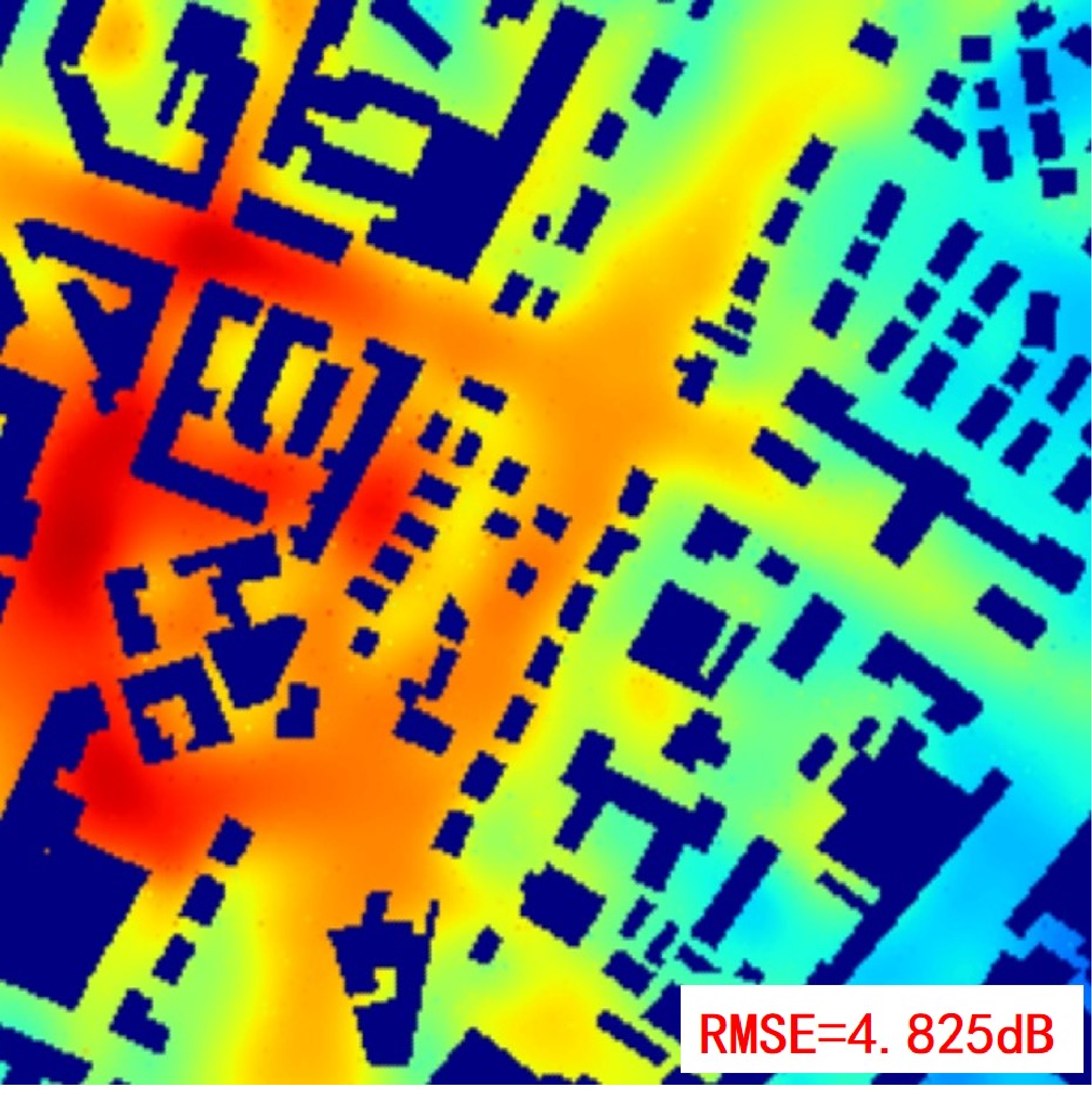} 
        \caption*{\small{SVGP}}
    \end{minipage}
    \hspace{-0.16cm}  
    \begin{minipage}[t]{0.027\textwidth}  
        \centering
        \includegraphics[width=\textwidth]{cbar.eps} 
    \end{minipage}
    \vspace{0.1cm}

    \begin{minipage}[t]{0.147\textwidth}  
        \centering
        \includegraphics[width=\textwidth]{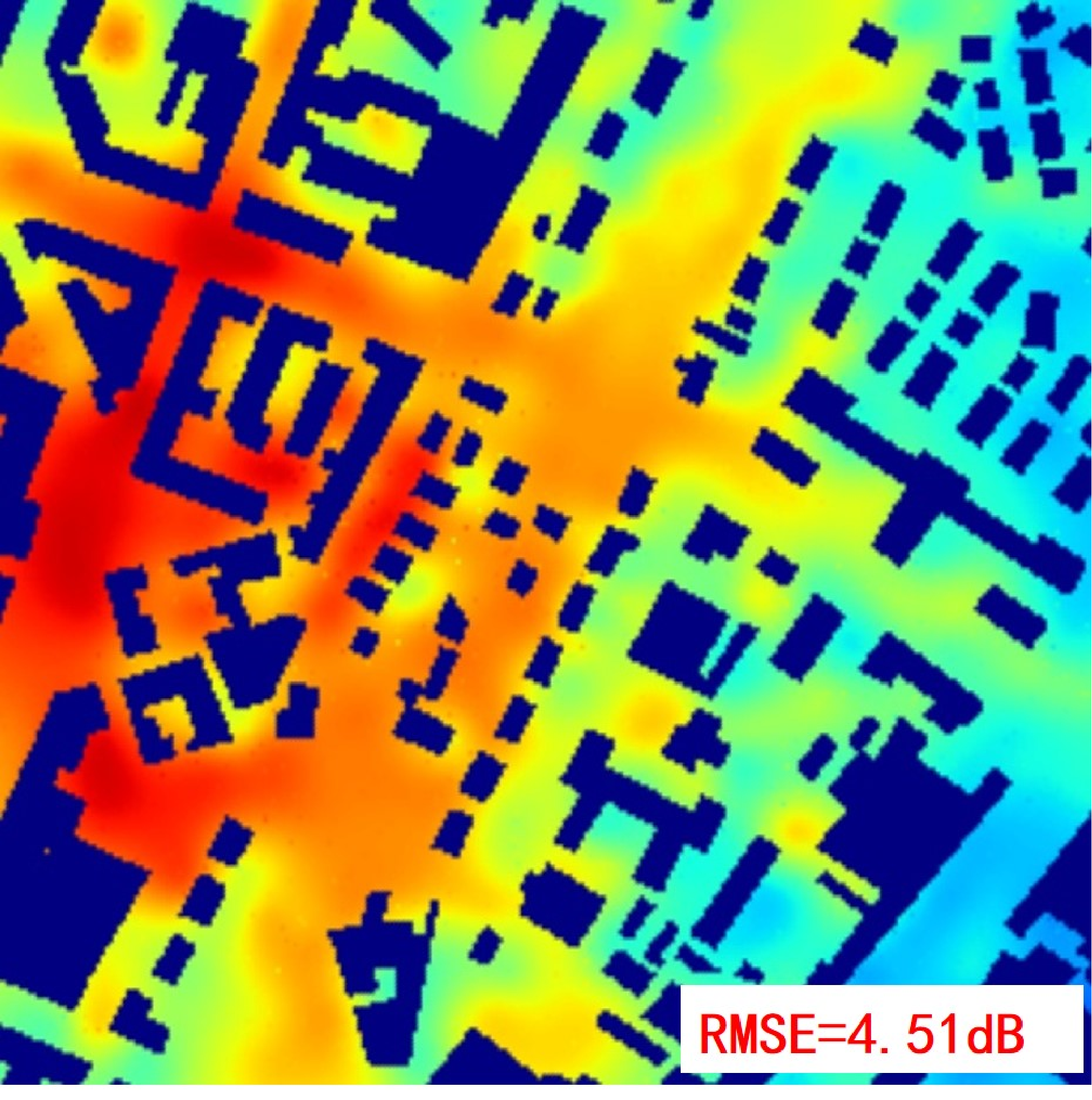} 
        \caption*{\small{M-OSVGP-GOIPS}}
    \end{minipage}
    \hfill
    \begin{minipage}[t]{0.147\textwidth}  
        \centering
        \includegraphics[width=\textwidth]{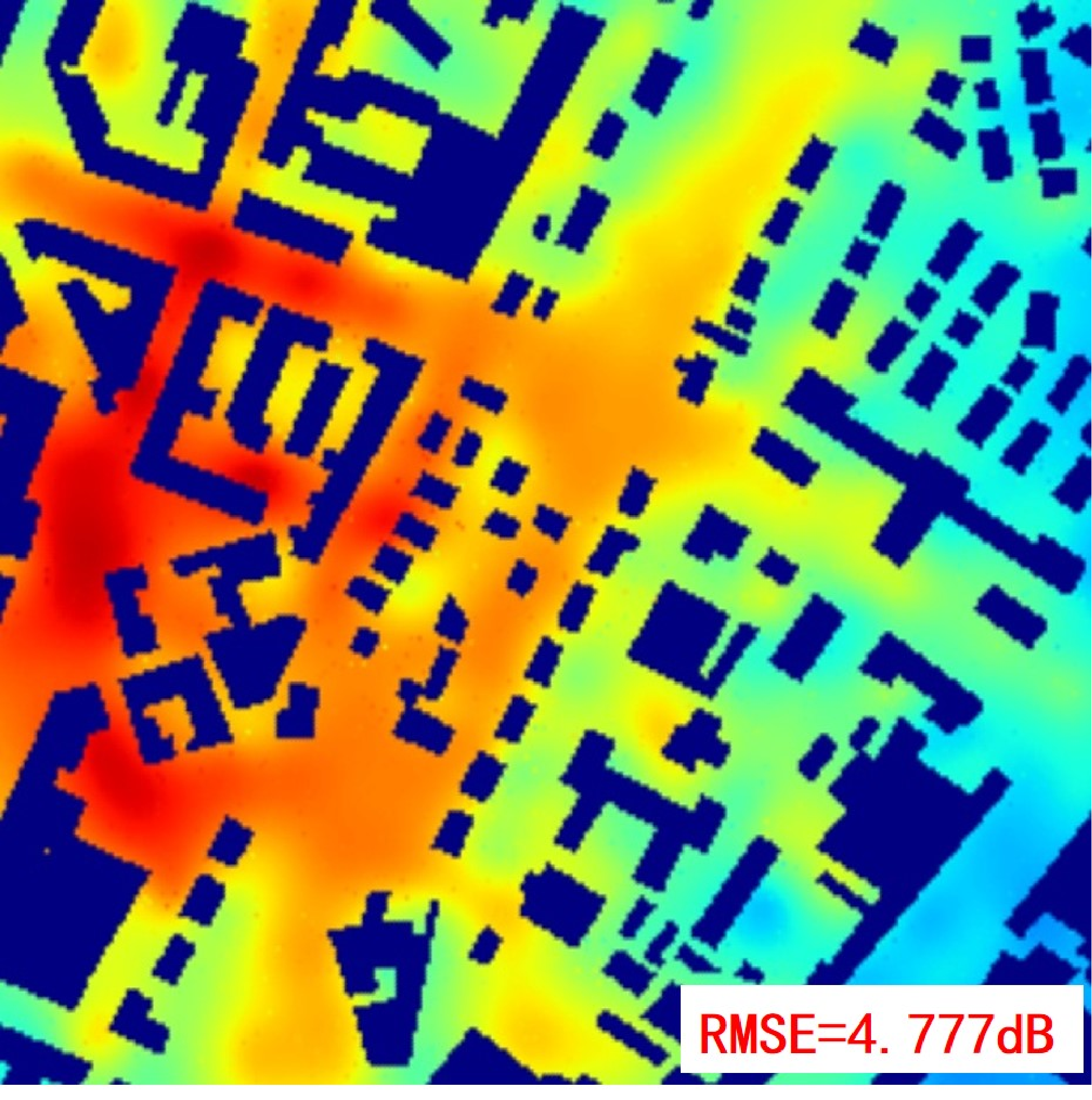} 
        \caption*{\small{SSVGP}}
    \end{minipage}
    \hfill
    \begin{minipage}[t]{0.147\textwidth}  
        \centering
        \includegraphics[width=\textwidth]{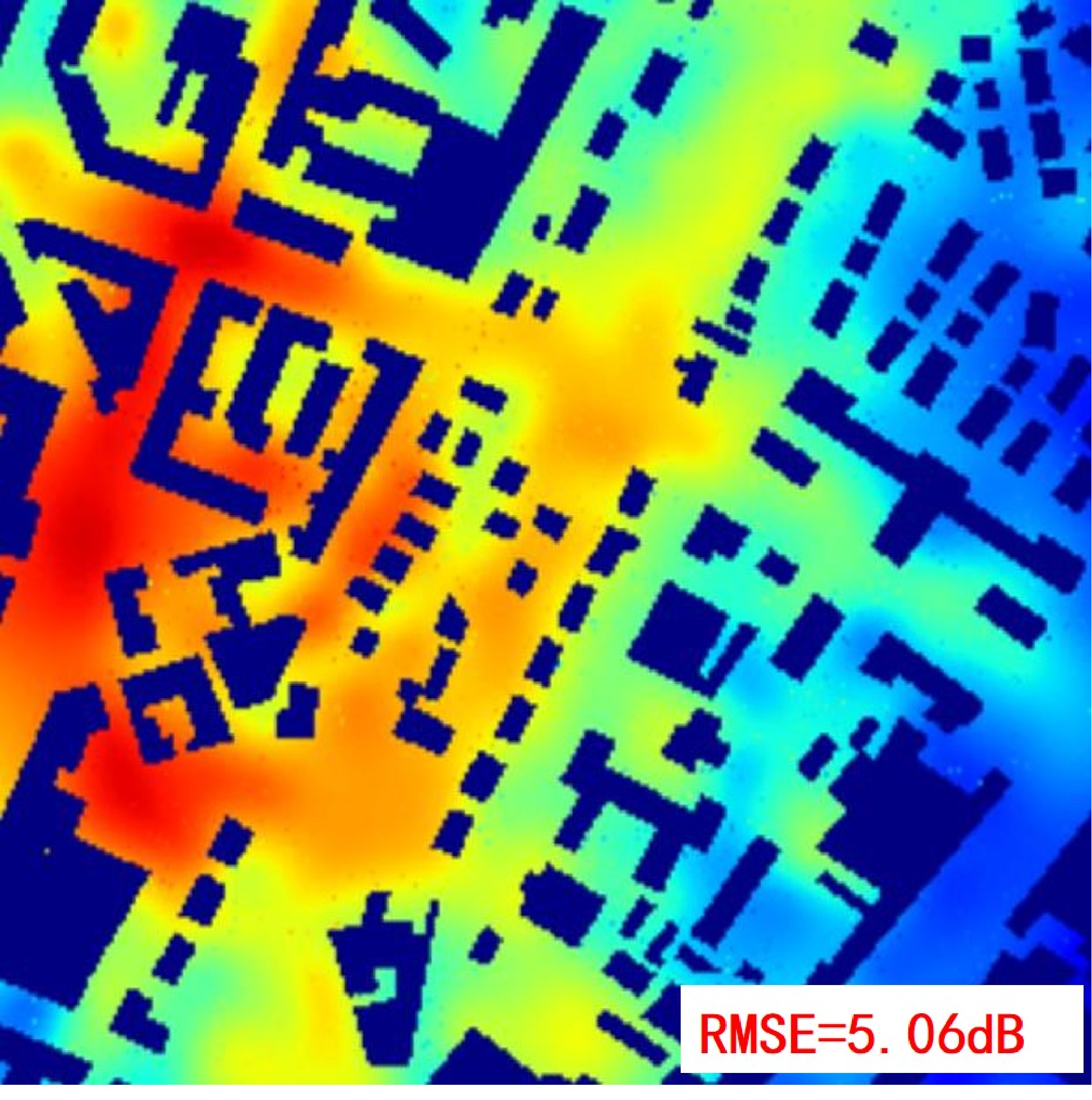} 
        \caption*{\small{Memory-DGP}}
    \end{minipage}
    \hspace{-0.16cm}  
    \begin{minipage}[t]{0.027\textwidth}  
        \centering
        \includegraphics[width=\textwidth]{cbar.eps} 
    \end{minipage}
    
    \caption{Illustrations of true map and reconstructed radio map for different methods after processing the fifth batch of measurements in the \textit{urban} scenario.}
    \label{fig:comparison}
\end{figure}

In Fig.~\ref{fig:comparison}, we illustrate the reconstructed radio maps obtained by our two proposed methods and the five baselines after processing the fifth batch of measurements in the \textit{urban} scenario. 
It can be seen that our proposed M-OSVGP and M-OSVGP-GOIPS achieve higher reconstruction accuracy, with reconstructed radio maps more closely matching the true radio map.
Compared with the baselines, our methods capture finer spatial variations and produce smoother interpolation in areas with sparse measurements.

\begin{figure*}[htbp]
    \centering
    \begin{subfigure}[b]{0.32\textwidth} 
        \centering
        \includegraphics[width=\textwidth]{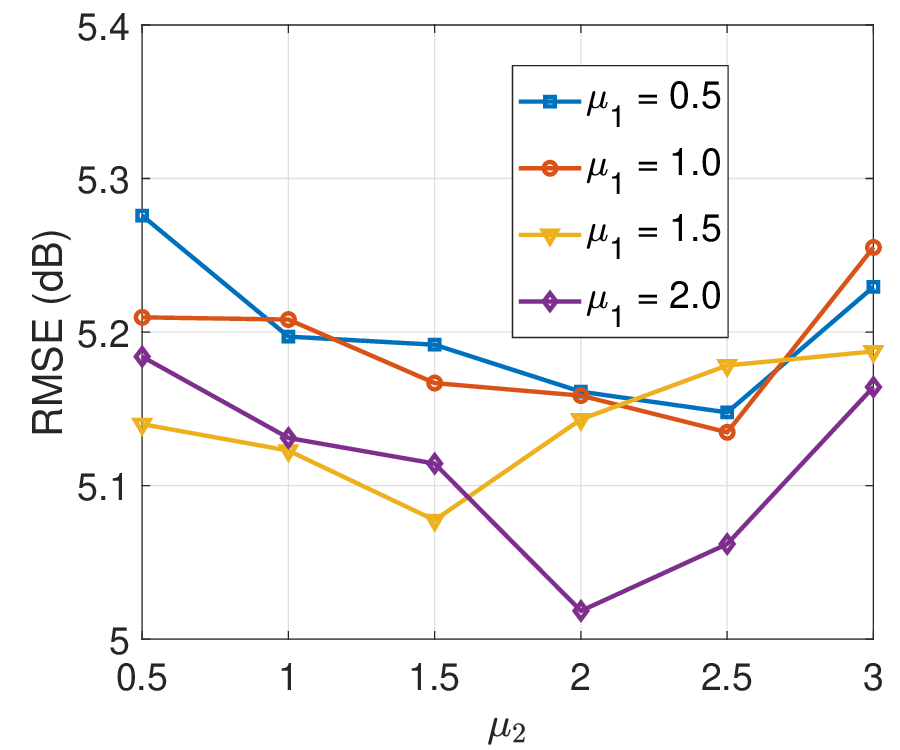}
        \caption{Batch 2}
        \label{fig:Batch=2}
    \end{subfigure}
    \hfill
    \begin{subfigure}[b]{0.32\textwidth} 
        \centering
        \includegraphics[width=\textwidth]{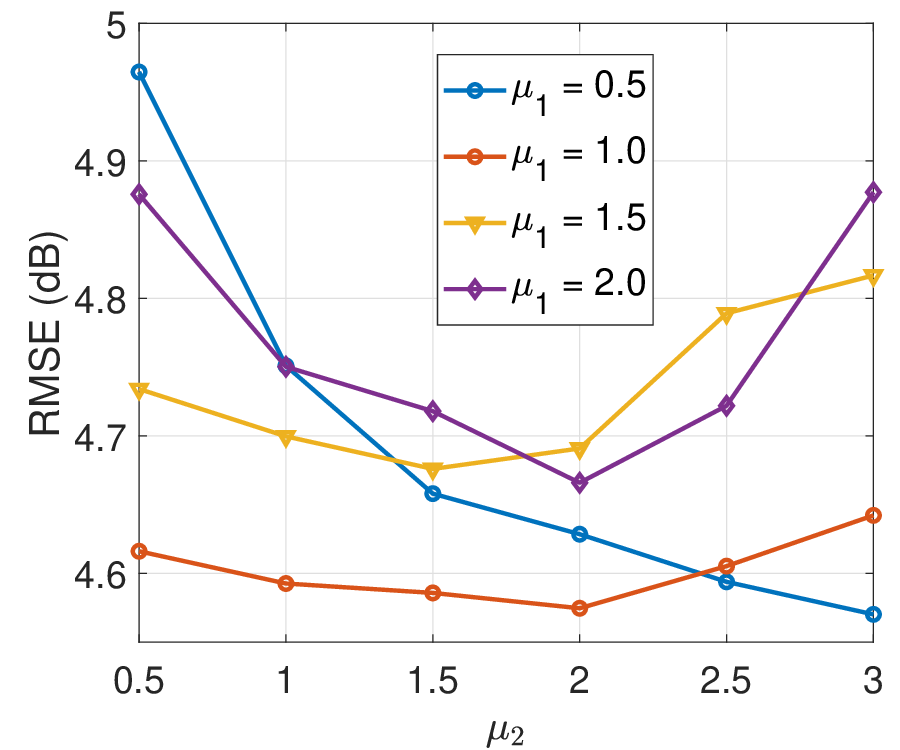}
        \caption{Batch 6}
        \label{fig:Batch=6}
    \end{subfigure}
    \hfill
    \begin{subfigure}[b]{0.32\textwidth} 
        \centering
        \includegraphics[width=\textwidth]{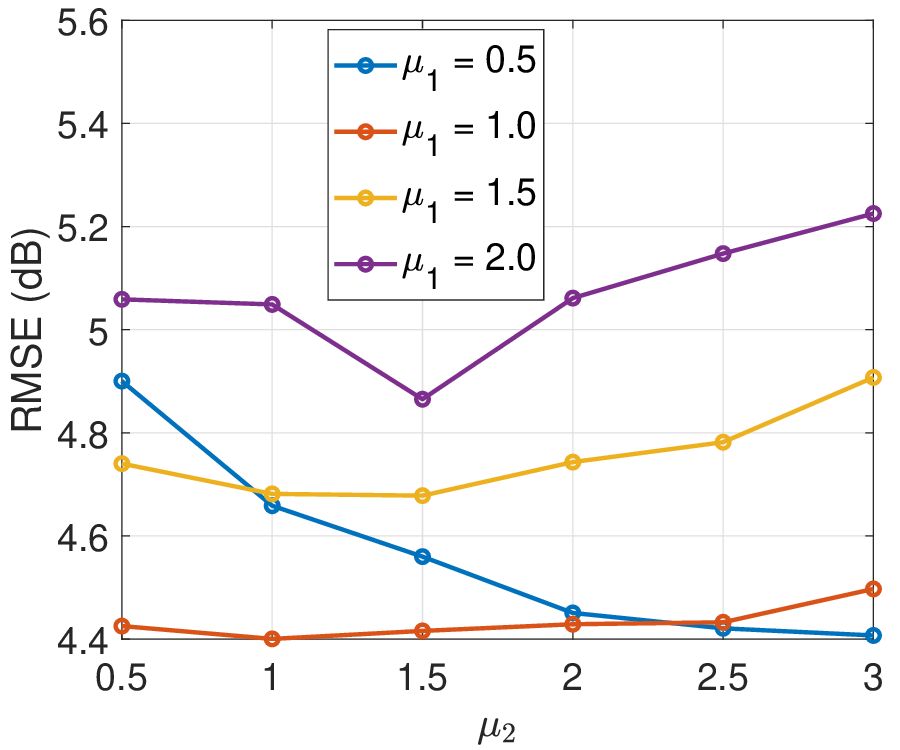}
        \caption{Batch 10}
        \label{fig:Batch=10}
    \end{subfigure}
    \caption{RMSE of M-OSVGP with different $\mu_1$ and $\mu_2$ after processing different measurement batches in the \textit{urban} scenario.}
    \label{fig:vary_mu1_mu2}
\end{figure*}

\subsection{Impact of Key Parameters}
Here, we study the impact of key parameters on the reconstruction accuracy of our proposed M-OSVGP method in the \textit{urban} scenario.\footnote{The results of M-OSVGP-GOIPS and the \textit{campus} scenario are similar and thus omitted.}

Fig.~\ref{fig:vary_mu1_mu2} illustrates the RMSE under different values of $\mu_1$ and $\mu_2$ in the hybrid objective \eqref{hybrid objective}, after processing the 2nd, 6th, and 10th measurement batches.
We see that for a given $\mu_1$, the RMSE curve with respect to $\mu_2$ generally exhibits a U-shape. 
This is because $\mu_2$ regulates the contributions of historical measurements during model update, as shown in \eqref{hybrid objective}.
When $\mu_2$ is very small, the model relies excessively on the latest batch, leading to unstable updates; when $\mu_2$ is very large, the updates become overly conservative and less adaptive. 
Similarly, as $\mu_1$ controls the influence of the previous posterior approximation in  \eqref{hybrid objective}, a small value may cause discontinuous updates that compromise spatial coherence, while a large value leads to conservative updates that under-utilize new spatial information.

Moreover, in Fig.~\ref{fig:vary_mu1_mu2}, we can see that the preferred values of $\mu_1$ and $\mu_2$ evolve as more batches are processed. 
At the early stage (e.g., Batch 2), when the number of measurements is limited and the model uncertainty remains high, relatively larger coefficients help stabilize the updates.
As more batches are accumulated, the model obtains more reliable spatial representations, and thus moderate values (e.g., $(\mu_1=1, \mu_2=2)$ at batch 6 and $(\mu_1=1, \mu_2=1)$ at batch 10) achieve the best performance.
This reflects a transition from relying more on historical knowledge for stability to allowing greater flexibility for adapting to new data.

\begin{figure}[t]
    \centering
    \begin{subfigure}[b]{0.24\textwidth}
        \centering
        \includegraphics[width=\textwidth]{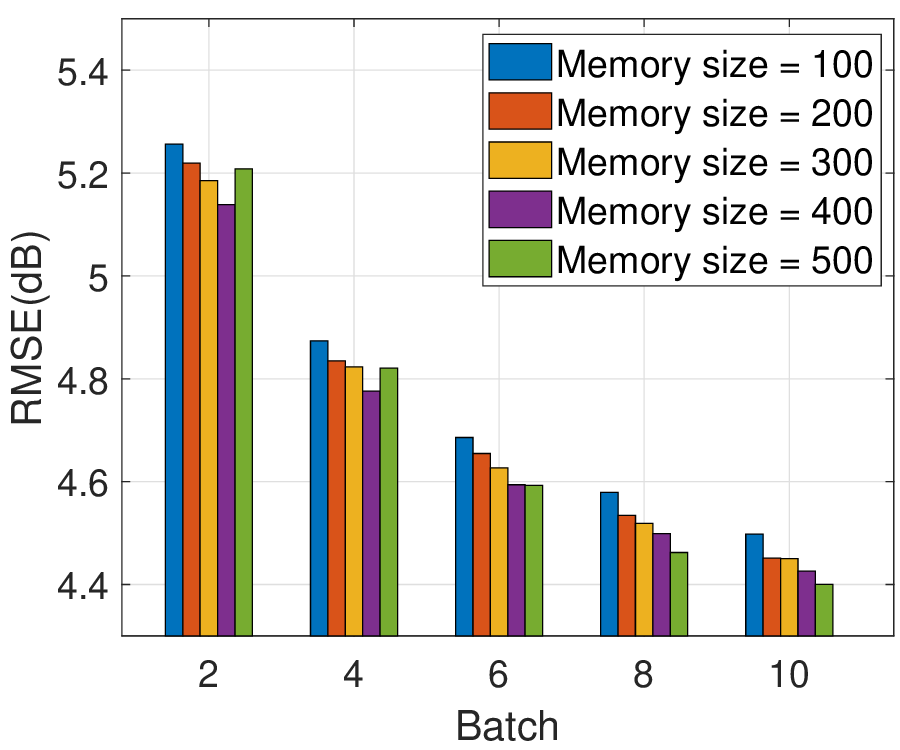}
        \caption{Memory size}
        \label{fig:RMSE_memorysize}
    \end{subfigure}
    \hfill
    \hspace{-0.61cm}  
    \begin{subfigure}[b]{0.24\textwidth}
        \centering
        \includegraphics[width=\textwidth]{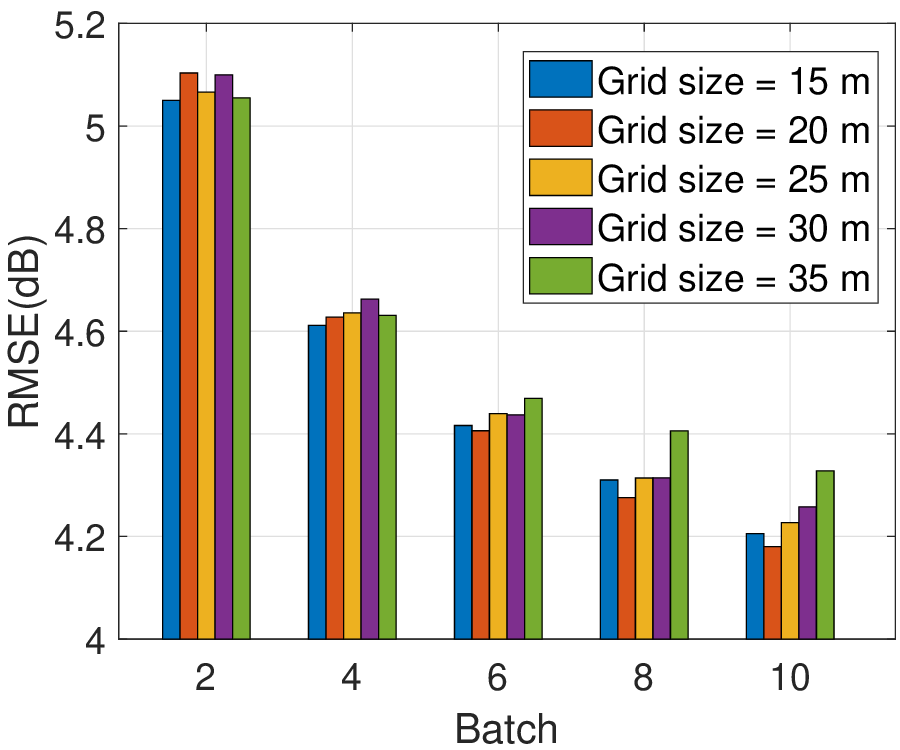}
        \caption{Grid size}
        \label{fig:RMSE_gridsize}
    \end{subfigure}
    \caption{RMSE of M-OSVGP under different memory sizes and grid sizes in the \textit{urban} scenario.}
    \label{fig: size}
\end{figure}

Next, in Fig.~\ref{fig: size}, we illustrate the RMSE performance under different memory sizes $N_\mathcal{M}$ in our memory subset mechanism (Section \ref{Hybrid Objective}) and grid sizes in GOIPS throughout the updating process. 
As shown in Fig.~\ref{fig:RMSE_memorysize}, during the early learning stage (Batches 2–4), moderate memory sizes (e.g., $N_\mathcal{M}=400$) achieve the best performance, as they offer sufficient spatial diversity without overwhelming the limited current measurements.
As the updating progresses (Batches 6–10), larger memory sizes become increasingly beneficial, with $N_\mathcal{M}=500$ outperforming smaller ones.
A larger memory can store a more spatially diverse set of historical measurements, which is crucial for mitigating catastrophic forgetting while accommodating new information.

From Fig.~\ref{fig:RMSE_gridsize}, we observe that, the RMSE  is largely insensitive to grid sizes in the early stage (Batches 2–4) since the model acquired sufficient knowledge of global spatial structure under sparse measurements.
As spatial knowledge accumulates (Batches 6–10), we see that intermediate grid sizes (about 20 m) yield the best results. 
This is because coarse grids (e.g., 30-35m) are too large to capture fine-grained geographical heterogeneity, while overly fine grids (15 m) introduce excessive spatial fragmentation, which undermines the kernel function’s ability to capture smooth correlations in radio propagation.

\begin{figure}[htbp]
    \centering
    \begin{subfigure}[b]{0.24\textwidth}
        \centering
        \includegraphics[width=\textwidth]{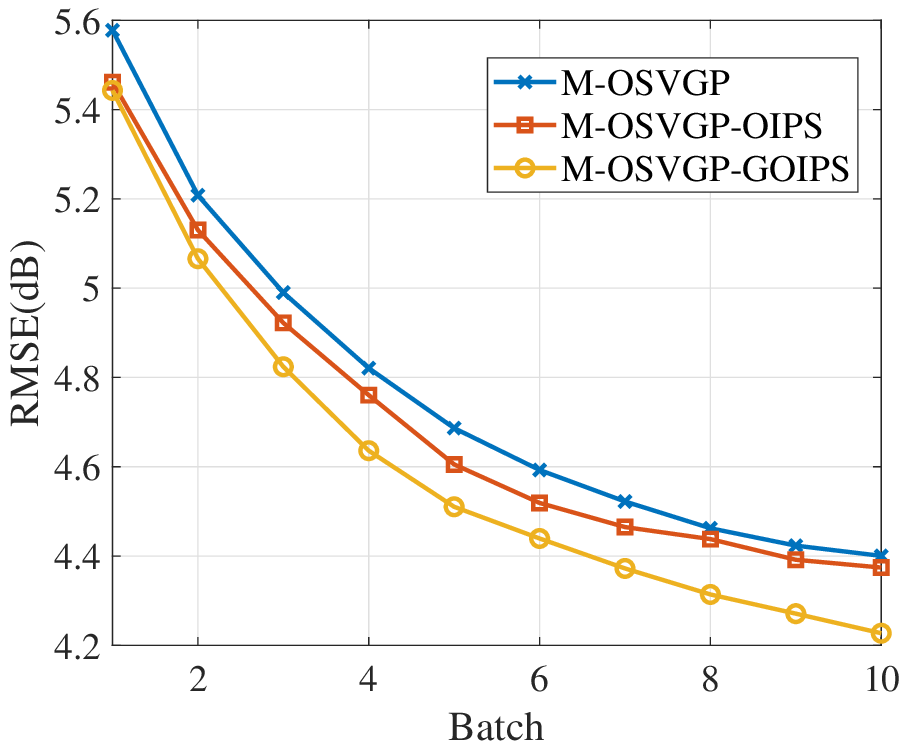}
        \caption{RMSE}
        \label{fig:RMSE_GOIPS_OIPS_Random}
    \end{subfigure}
    \hfill
        \begin{subfigure}[b]{0.24\textwidth}
        \centering
        \includegraphics[width=\textwidth]{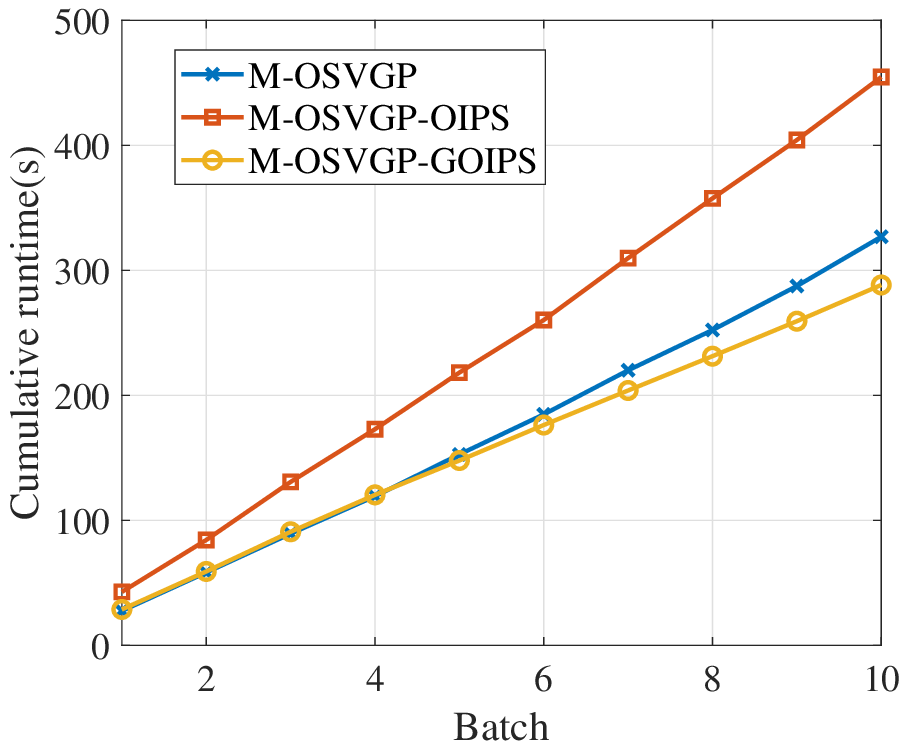}
        \caption{Cumulative runtime}
        \label{fig:Time_GOIPS_OIPS_Random}
    \end{subfigure}
    \caption{RMSE and cumulative runtime of three M-OSVGP variants under different inducing point initialization strategies in the \textit{urban} scenario.}
    \label{fig: GOIPS_OIPS_Random}
\end{figure}

\subsection{Effectiveness of GOIPS}
Finally, to validate the effectiveness of our proposed inducing point selection method, GOIPS, compare the RMSE and cumulative runtime of three M-OSVGP variants under different inducing point initialization strategies, as shown in Fig.~\ref{fig: GOIPS_OIPS_Random}.
The first is the original M-OSVGP with its default random initialization.
The second is M-OSVGP-GOIPS, which replaces random initialization with GOIPS.
The third, M-OSVGP-OIPS, replaces random initialization with the Online Inducing Point Selection (OIPS) algorithm \cite{GalyFajou2021}, which sequentially adds new inducing points  based solely on a kernel similarity threshold. 
We can see that  GOIPS achieves the lowest RMSE throughout the updating process while requiring significantly less cumulative runtime than the other two variants.
This is because GOIPS can adaptively adjust the inducing set size and select representative points according to current measurement patterns and spatial distribution, and thus maintains a favorable trade-off between computational cost and approximation accuracy.

\begin{figure}[htbp]
    \centering
    \begin{subfigure}[b]{0.24\textwidth}
        \centering
        \includegraphics[width=\textwidth]{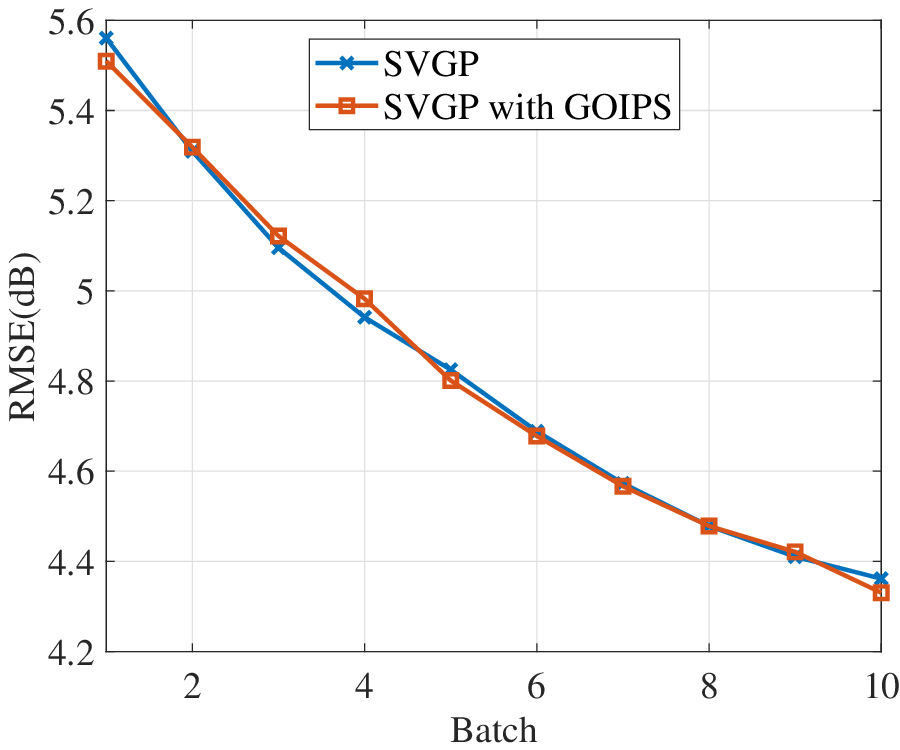}
        \caption{RMSE}
        \label{fig:RMSE_SVGP_GOIPS}
    \end{subfigure}
    \hfill
        \begin{subfigure}[b]{0.24\textwidth}
        \centering
        \includegraphics[width=\textwidth]{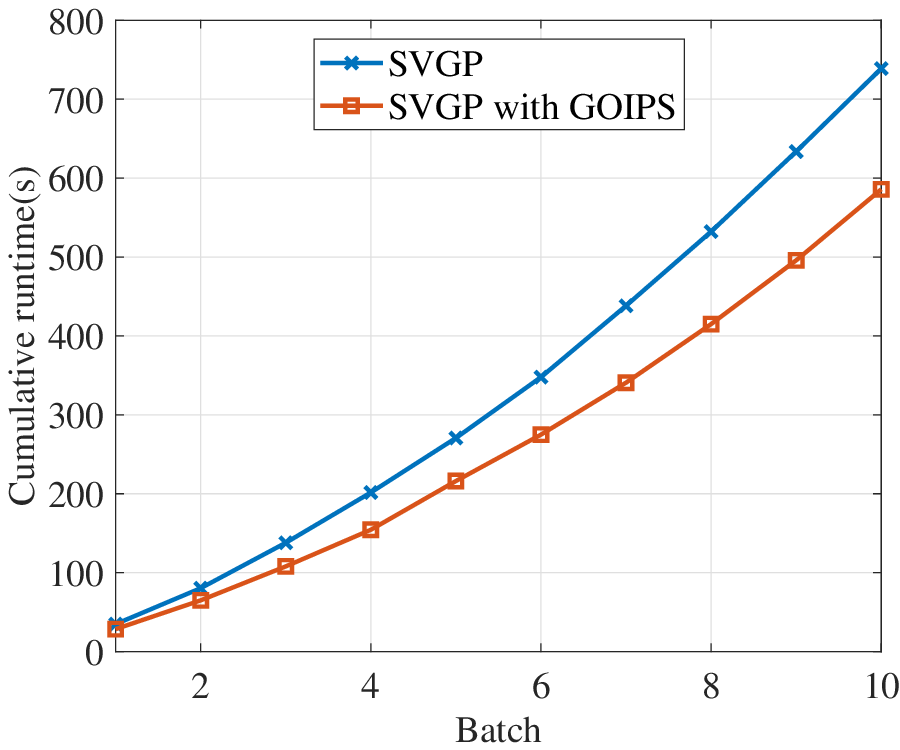}
        \caption{Cumulative runtime}
        \label{fig:Time_SVGP_GOIPS}
    \end{subfigure}
 \caption{RMSE and cumulative runtime of SVGP and SVGP with GOIPS in the \textit{urban} scenario.}
    \label{fig: SVGP_GOIPS}
\end{figure}

\begin{figure}[htbp]
    \centering
    \begin{subfigure}[b]{0.24\textwidth}
        \centering
        \includegraphics[width=\textwidth]{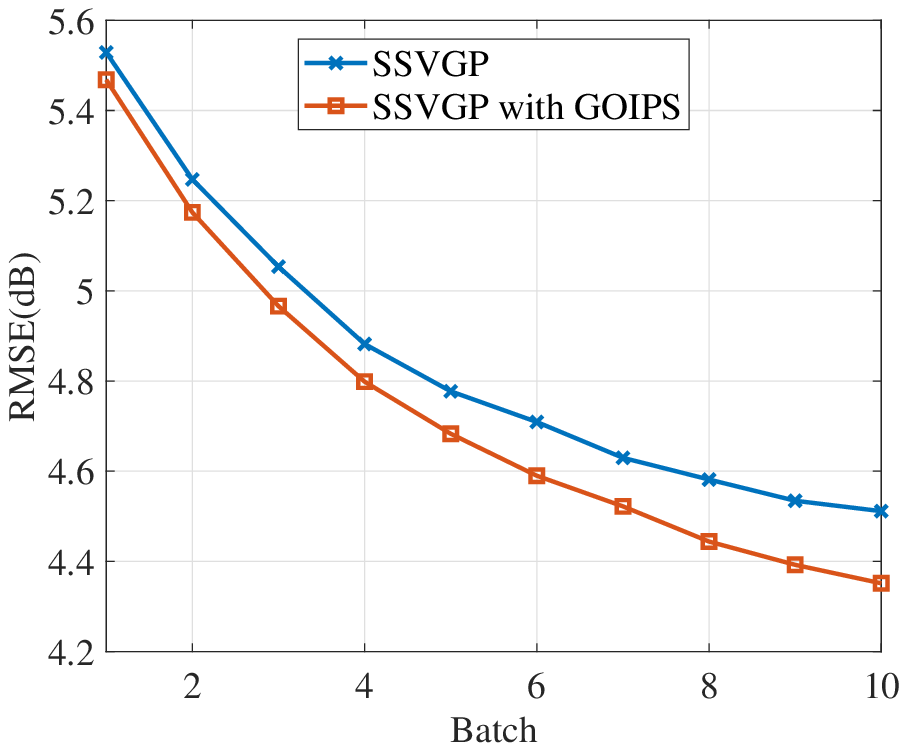}
        \caption{RMSE}
        \label{fig:RMSE_SSVGP_GOIPS}
    \end{subfigure}
    \hfill
    \begin{subfigure}[b]{0.24\textwidth}
        \centering
        \includegraphics[width=\textwidth]{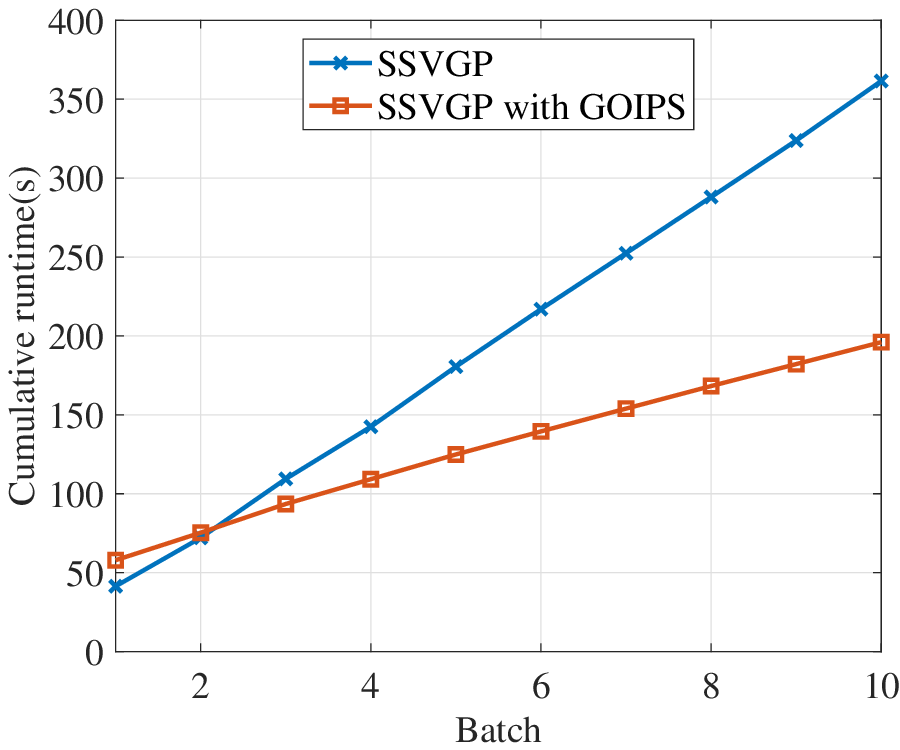}
        \caption{Cumulative runtime}
        \label{fig:Time_SSVGP_GOIPS}
    \end{subfigure}
 \caption{RMSE and cumulative runtime of SSVGP and SSVGP with GOIPS in the \textit{urban} scenario.}
    \label{fig: SSVGP_GOIPS}
\end{figure}

To further verify the general applicability of GOIPS, we then integrate it as a standalone module into two standard sparse GP models: the batch SVGP and the online SSVGP, and compare their performance in terms of RMSE and cumulative runtime. Fig.~\ref{fig: SVGP_GOIPS} shows that, when integrated into SVGP, GOIPS achieves substantial computational savings while preserving almost identical  reconstruction accuracy. In Fig.~\ref{fig: SSVGP_GOIPS}, the integration with SSVGP brings both computational efficiency gains and clear accuracy improvements. These results confirm that GOIPS provides high-quality representative inducing points at lower cost, making it an effective plug-in component  across different GP frameworks.

\section{Conclusion}
In this paper, we have proposed a memory-based online sparse variational Gaussian process framework, M-OSVGP, for adaptive and efficient radio map reconstruction from streaming spectrum measurements. 
By employing a hybrid objective together with a representative memory subset, our method incrementally incorporates new measurement data while mitigating catastrophic forgetting and maintaining a constant computational complexity. 
To further improve posterior approximation, we have introduced the GOIPS algorithm, which dynamically adjusts both the locations and the number of inducing points according to measurement distribution and spatial correlation. 
Extensive simulations demonstrate that our proposed M-OSVGP and M-OSVGP-GOIPS consistently achieve superior reconstruction accuracy and computational efficiency compared with existing batch and online baselines.
Moreover, the results show that  GOIPS not only achieves lower RMSE and runtime compared to other selection strategies, but also significantly improves standard SVGP and SSVGP when integrated as a plug-in component.


\bibliographystyle{IEEEtran}
\bibliography{radiomap_updating}
\appendix
\section{Appendix}
\subsection{Derivation of the Online ELBO in \eqref{online ELBO}}\label{Online ELBO Derivation}
By using the factorized form of the variational distributions and applying Bayes' theorem, we can obtain 
\begin{align*}
    \frac{p(\mathbf{f}|\theta_{n-1})}{q^{n-1}(\mathbf{f})}&= \frac{p(\mathbf{f}_{\ne \mathbf{u}_{n-1}}|\mathbf{u}_{n-1},\theta_{n-1})p(\mathbf{u}_{n-1}|\theta_{n-1})}{p(\mathbf{f}_{\ne \mathbf{u}_{n-1}}|\mathbf{u}_{n-1},\theta_{n-1})q^{n-1}(\mathbf{u}_{n-1})} \nonumber\\&= \frac{p(\mathbf{u}_{n-1}|\theta_{n-1})}{q^{n-1}(\mathbf{u}_{n-1})},\\
    \frac{q^{n}(\mathbf{f})}{p(\mathbf{f}|\theta_{n})} &= \frac{p(\mathbf{f}|\mathbf{u}_{n},\theta_{n})q^{n}(\mathbf{u}_{n})}{p(\mathbf{f}|\mathbf{u}_{n},\theta_{n})p(\mathbf{u}_{n}|\theta_n)}=\frac{q^{n}(\mathbf{u}_{n})}{p(\mathbf{u}_{n}|\theta_n)}.
\end{align*}
Then, we can rewrite the last term in \eqref{KL divergence}, which constitutes the ELBO, as follows:
\begin{align}
    &\mathcal{L}_\text{online}\nonumber\\&= \int q^n(\mathbf{f}) \log \frac{q^{n}(\mathbf{u}_{n})p(\mathbf{u}_{n-1}|\theta_{n-1})}{p(\mathbf{u}_{n}|\theta_n)p(\boldsymbol{\psi}_{n}| \mathbf{f})q^{n-1}(\mathbf{u}_{n-1})}d\mathbf{f} \label{ELBO_1}\\
    &= \int p(\mathbf{f}_{\ne \mathbf{u}_{n}}|\mathbf{u}_{n},\theta_{n})q^{n}(\mathbf{u}_{n}) \log \frac{q^{n}(\mathbf{u}_{n})p(\mathbf{u}_{n-1}|\theta_{n-1})}{p(\mathbf{u}_{n}|\theta_n)q^{n-1}(\mathbf{u}_{n-1})}d\mathbf{f}_{\ne \mathbf{u}_{n}}d\mathbf{u}_{n} \nonumber\\&\phantom{-}- \int q^n(\mathbf{f}) \log p(\boldsymbol{\psi}_{n}| \mathbf{f})d\mathbf{f} \nonumber\\
    &= \int q^{n}(\mathbf{u}_{n}) \log \frac{q^{n}(\mathbf{u}_{n})}{p(\mathbf{u}_{n}|\theta_n)}d\mathbf{u}_{n}  - \int q^n(\mathbf{f}) \log p(\boldsymbol{\psi}_{n}| \mathbf{f})d\mathbf{f}  \nonumber\\&\phantom{-} + \int q^{n}(\mathbf{u}_{n}) \log \frac{p(\mathbf{u}_{n-1}|\theta_{n-1})}{q^{n-1}(\mathbf{u}_{n-1})}d\mathbf{u}_{n}\nonumber\\
    &= \text{KL}[q^{n}(\mathbf{u}_{n})|p(\mathbf{u}_{n}|\theta_n)] - \int q^n(\mathbf{f}) \log p(\boldsymbol{\psi}_{n}| \mathbf{f})d\mathbf{f} \nonumber\\
    &\phantom{-}+ \int q^{n}(\mathbf{u}_{n}) \log\left[\frac{p(\mathbf{u}_{n-1}|\theta_{n-1})}{q^{n-1}(\mathbf{u}_{n-1})}\frac{q^{n}(\mathbf{u}_{n-1})}{q^{n}(\mathbf{u}_{n-1})}\right]d\mathbf{u}_{n}\nonumber\\
    &= \text{KL}[q^{n}(\mathbf{u}_{n})|p(\mathbf{u}_{n}|\theta_n)]-\sum_{\psi_i\in \boldsymbol{\psi}_n}\mathbb{E}_{q^n_{\mathbf{u}}(f_i)}[\log p(\psi_i|f_i)] \nonumber\\
    & \phantom{-} + \text{KL}[q^{n}(\mathbf{u}_{n-1})|q^{n-1}(\mathbf{u}_{n-1})] \nonumber\\
    &\phantom{-} - \text{KL}[q^{n}(\mathbf{u}_{n-1})|p(\mathbf{u}_{n-1}|\theta_{n-1})].
\end{align}
This completes the derivation of the online ELBO in \eqref{online ELBO}.

\subsection{Derivation of the Optimal Form of \texorpdfstring{$q_{\text{opt}}^n(\mathbf{u}_n)$}{q_n} in \eqref{optimal q(u)} }
\label{optimal form of q^n(u_n)}
We first present the hybrid objective $\mathcal{L}_\text{hybrid}$ in \eqref{hybrid objective} into the integral form:
\begin{align}\label{L_hybrid}
    &\mathcal{L}_\text{hybrid}\nonumber\\&=\int q^n(\mathbf{f})\log \frac{q^n(\mathbf{u}_n)}{p(\mathbf{u}_{n}|\theta_n)}d\mathbf{f} - \int q^n(\mathbf{f}) \log p(\boldsymbol{\psi}_{n}| \mathbf{f})d\mathbf{f} \nonumber\\
    &\phantom{-}+ \mu_1 \int q^n(\mathbf{f})\log \frac{p(\mathbf{u}_{n-1}|\theta_{n-1})}{q^{n-1}(\mathbf{u}_{n-1})}d\mathbf{f} \nonumber \\& \phantom{-}- \mu_2 \int q^n(\mathbf{f}) \log p(\boldsymbol{\psi}_{\mathcal{M}}| \mathbf{f})d\mathbf{f}.
\end{align}
By taking its derivative of the hybrid objective $\mathcal{L}_\text{hybrid}$ and a Lagrange term that normalizes $q^n(\mathbf{u}_n)$ with respect to $q^n(\mathbf{u}_n)$ and letting it be $0$, we have 
\begin{align}
    &\frac{d\mathcal{L}_\text{hybrid}}{dq^n(\mathbf{u}_n)} + \lambda \nonumber\\
    &=\log \frac{q^n(\mathbf{u}_n)}{p(\mathbf{u}_n|\theta_n)}+ 1 -\int p(\mathbf{f}_{\ne \mathbf{u}_{n}}|\mathbf{u}_n)\log p(\boldsymbol{\psi}_n|\mathbf{f})d\mathbf{f}_{\ne \mathbf{u}_{n}}\nonumber\\
    &\phantom{-}+ \mu_1 \int p(\mathbf{u}_{n-1}|\mathbf{u}_n)\log \frac{p(\mathbf{u}_{n-1}|\theta_{n-1})}{q^{n-1}(\mathbf{u}_{n-1})}d\mathbf{u}_{n-1} \nonumber\\
    & \phantom{-} - \mu_2 \int p(\mathbf{f}_{\ne \mathbf{u}_{n}}|\mathbf{u}_n)\log p(\boldsymbol{\psi}_\mathcal{M}|\mathbf{f})d\mathbf{f}_{\ne \mathbf{u}_{n}} + \lambda =0
\end{align}
Then, the optimal variational distribution $q_\text{opt}^n(\mathbf{u}_n)$ can be derived as:
\begin{align}\label{optimal q(u) appendix}
    q_\text{opt}^n(\mathbf{u}_n) &= \frac{1}{C}p(\mathbf{u}_n|\theta_n)\exp\Big( \int p(\mathbf{f}|\mathbf{u}_n)\log p(\boldsymbol{\psi}_n|\mathbf{f}) d\mathbf{f} \nonumber\\
    &\phantom{-}+ \mu_1\int p(\mathbf{u}_{n-1}|\mathbf{u}_n)\log\frac{q^{n-1}(\mathbf{u}_{n-1})}{p(\mathbf{u}_{n-1}|\theta_{n-1})}d\mathbf{u}_{n-1} \nonumber\\
    &\phantom{-} + \mu_2 \int p(\mathbf{f}_m|\mathbf{u}_n)\log p(\boldsymbol{\psi}_\mathcal{M}|\mathbf{f}_m) d\mathbf{f}_m \Big).
\end{align}
where $C$ is a constant and $\log C=\lambda+1$. 
Here, we have dropped $\theta_n$ from $p(\mathbf{u}_n|\theta_n)$ to lighten the notation.
Note that 
\begin{align}
    q^{n-1}(\mathbf{u}_{n-1})&=\mathcal{N}(\mathbf{u}_{n-1};\mathbf{m}_{\mathbf{u}_{n-1}},\mathbf{S}_{\mathbf{u}_{n-1}}),\nonumber\\
    p(\mathbf{u}_{n-1}|\theta_{n-1})&=\mathcal{N}(\mathbf{u}_{n-1};0, \mathbf{K}_{\mathbf{u}_{n-1}\mathbf{u}_{n-1}}), \nonumber
\end{align}
and define 
\begin{align}
    \mathbf{Q}_{\mathbf{f}} &= \mathbf{K_{ff}} - \mathbf{K}_{\mathbf{f}\mathbf{u}_n} \mathbf{K}_{\mathbf{u}_n\mathbf{u}_n}^{-1} \mathbf{K}_{\mathbf{u}_n\mathbf{f}}, \nonumber\\
    \mathbf{Q}_{\mathbf{f}_m} &= \mathbf{K}_{\mathbf{f}_m\mathbf{f}_m} - \mathbf{K}_{\mathbf{f}_m\mathbf{u}_n} \mathbf{K}_{\mathbf{u}_n\mathbf{u}_n}^{-1} \mathbf{K}_{\mathbf{u}_n\mathbf{f}_m}, \nonumber\\
    \mathbf{Q}_{\mathbf{u}_{n-1}} &= \mathbf{K}_{\mathbf{u}_{n-1}\mathbf{u}_{n-1}} - \mathbf{K}_{\mathbf{u}_{n-1}\mathbf{u}_n} \mathbf{K}_{\mathbf{u}_n\mathbf{u}_n}^{-1} \mathbf{K}_{\mathbf{u}_n\mathbf{u}_{n-1}}, \nonumber\\
    \mathbf{D}_{\mathbf{u}_{n-1}}&=(\mathbf{S}_{\mathbf{u}_{n-1}}^{-1}-\mathbf{K}_{\mathbf{u}_{n-1}\mathbf{u}_{n-1}}^{-1})^{-1}. \nonumber
\end{align}
Then, we can write each term in the exponents in \eqref{optimal q(u) appendix} as follows:
\begin{align}
    \text{E}_1 &= \int p(\mathbf{f}|\mathbf{u}_n)\log p(\boldsymbol{\psi}_n|\mathbf{f}) d\mathbf{f}\nonumber\\
    & = \int \mathcal{N}(\mathbf{f};\mathbf{K}_{\mathbf{fu}_n} \mathbf{K}_{\mathbf{u}_n\mathbf{u}_n}^{-1}\mathbf{u}_n,\mathbf{Q_f}) \log \mathcal{N}(\boldsymbol{\psi}_n;\mathbf{f}, \sigma_{\psi}^2\mathbf{I}_{N_n})\nonumber\\
    &= \log \mathcal{N}(\boldsymbol{\psi}_n;\mathbf{K}_{\mathbf{fu}_n} \mathbf{K}_{\mathbf{u}_n\mathbf{u}_n}^{-1}\mathbf{u}_n, \sigma_{\psi}^2\mathbf{I}_{N_n}) +\Delta_1, \\
    \text{E}_2 &= \int p(\mathbf{u}_{n-1}|\mathbf{u}_n)\log\frac{q^{n-1}(\mathbf{u}_{n-1})}{p(\mathbf{u}_{n-1}|\theta_{n-1})}d\mathbf{u}_{n-1} \nonumber\\
    & = \frac{1}{2}\int \mathcal{N}(\mathbf{u}_{n-1};\mathbf{K}_{\mathbf{u}_{n-1}\mathbf{u}_{n}} \mathbf{K}_{\mathbf{u}_n\mathbf{u}_n}^{-1}\mathbf{u}_n,\mathbf{Q}_{\mathbf{u}_{n-1}})\nonumber\\ &\phantom{-}\times \bigg(\log \frac{|\mathbf{K}_{\mathbf{u}_{n-1}\mathbf{u}_{n-1}}|}{|\mathbf{S}_{\mathbf{u}_{n-1}}|} -(\mathbf{u}_{n-1} - \mathbf{m}_{\mathbf{u}_{n-1}})^\top \nonumber \\& \phantom{-} \times \mathbf{S}_{\mathbf{u}_{n-1}}^{-1} (\mathbf{u}_{n-1} - \mathbf{m}_{\mathbf{u}_{n-1}}) + \mathbf{u}_{n-1}^\top \mathbf{K}_{\mathbf{u}_{n-1}\mathbf{u}_{n-1}}^{-1} \mathbf{u}_{n-1} \bigg) \nonumber\\
    & = \log \mathcal{N}(\mathbf{D}_{\mathbf{u}_{n-1}} \mathbf{S}_{\mathbf{u}_{n-1}}^{-1}\mathbf{m}_{\mathbf{u}_{n-1}}; \mathbf{K}_{\mathbf{u}_{n-1}\mathbf{u}_{n}} \mathbf{K}_{\mathbf{u}_n\mathbf{u}_n}^{-1}\mathbf{u}_n, \mathbf{D}_{\mathbf{u}_{n-1}}) \nonumber\\ & \phantom{-} + \Delta_2, \\
    \text{E}_3 &= \int p(\mathbf{f}_m|\mathbf{u}_n)\log p(\boldsymbol{\psi}_{\mathcal{M}}|\mathbf{f}_m) d\mathbf{f}_m\nonumber\\
    & = \int \mathcal{N}(\mathbf{f}_m;\mathbf{K}_{\mathbf{f}_m\mathbf{u}_n} \mathbf{K}_{\mathbf{u}_n\mathbf{u}_n}^{-1}\mathbf{u}_n,\mathbf{Q}_{\mathbf{f}_m}) \log \mathcal{N}(\boldsymbol{\psi}_n;\mathbf{f}_m, \sigma_{\psi}^2\mathbf{I}_{N_\mathcal{M}})\nonumber\\
    &= \log \mathcal{N}(\boldsymbol{\psi}_\mathcal{M};\mathbf{K}_{\mathbf{f}_m\mathbf{u}_n} \mathbf{K}_{\mathbf{u}_n\mathbf{u}_n}^{-1}\mathbf{u}_n, \sigma_{\psi}^2\mathbf{I}_{N_\mathcal{M}}) +\Delta_3,
\end{align}
where  $\Delta_1, \Delta_2$, and $\Delta_3$ are given in \eqref{L_theta}.
By substituting these results back into \eqref{optimal q(u) appendix}, we can obtain the optimal variational distribution given in \eqref{optimal q(u)}.

\subsection{Implementation of the Hybrid Objective \texorpdfstring{$\mathcal{L}_{\text{hybrid}}$}{L(θ)}} \label{L calculate}
To calculate the hybrid objective under the optimal distribution, 
we first take the logarithm of the optimal distribution $q_{opt}^n(\mathbf{u}_n)$ in \eqref{optimal q(u) appendix} and rearrange it to obtain $\log \frac{q_{opt}^n(\mathbf{u}_n)}{p(\mathbf{u}_n|\theta_n)}$.
By substituting this expression into the hybrid objective in \eqref{L_hybrid}, we obtain
$\mathcal{L}_\text{hybrid}(q_{opt}^n(\mathbf{u}_n)) =-\log C$.
Thus, computing the hybrid objective reduces to calculating $\log C$. This can be obtained by 
integrating both sides of \eqref{optimal q(u) appendix} and then taking the logarithm, which yields:
\begin{align}\label{L_calculate}
    &\mathcal{L}_\text{hybrid} = \log\mathcal{N}(\hat{\boldsymbol{\psi}};0, \mathbf{K}_{\mathbf{\hat{f}}\mathbf{u}_n}\mathbf{K}_{\mathbf{u}_n\mathbf{u}_n}^{-1}\mathbf{K}_{\mathbf{u}_n\mathbf{\hat{f}}}+ \Sigma_{\hat{\boldsymbol{\psi}}}) + \Delta_1 + \mu_1 \Delta_2 \nonumber\\& + \frac{M}{2}(1-\mu_1)\log(2\pi) - \frac{M}{2}\log(\mu_1) + \frac{1}{2}(1-\mu_1)\log|\mathbf{D}_{\mathbf{u}_{n-1}}| \nonumber\\& +\mu_2 \Delta_3 + \frac{N_\mathcal{M}}{2}(1-\mu_2)\log(2\pi\sigma_{\psi}^2) - \frac{N_\mathcal{M}}{2}\log(\mu_2).
\end{align}
Let $\mathbf{S}_\psi = \mathbf{K}_{\mathbf{\hat{f}}\mathbf{u}_n}\mathbf{K}_{\mathbf{u}_n\mathbf{u}_n}^{-1}\mathbf{K}_{\mathbf{u}_n\mathbf{\hat{f}}}+ \Sigma_{\hat{\boldsymbol{\psi}}}$ and we rewrite the first term in \eqref{L_calculate} as:
\begin{align}\label{L_hybrid_1}
    &\mathcal{L}_{\text{hybrid}}^1 = \log\mathcal{N}(\hat{\boldsymbol{\psi}};0, \mathbf{S}_\psi) \\
    & = -\frac{N_n+N_\mathcal{M}+M}{2}\log(2\pi) -\frac{1}{2}\log\big|\mathbf{S}_\psi\big|
     -\frac{1}{2}\hat{\boldsymbol{\psi}}^\top\mathbf{S}_\psi^{-1}\hat{\boldsymbol{\psi}}.\nonumber
\end{align}
Since $\mathbf{S}_\psi$ is $(N_n+N_\mathcal{M}+M)$-dimensional, computing its determinant and inverse directly in \eqref{L_hybrid_1} is expensive and numerically unstable.
To address this, we provide the practical implementations of \eqref{L_calculate}. 
First, by using the matrix determinant lemma, we can compute the determinant of $\mathbf{S}_\psi$ as:
\begin{align}\label{log_det_Spsi}
    \log|\mathbf{S}_\psi|&=\log|\mathbf{K}_{\mathbf{\hat{f}}\mathbf{u}_n}\mathbf{K}_{\mathbf{u}_n\mathbf{u}_n}^{-1}\mathbf{K}_{\mathbf{u}_n\mathbf{\hat{f}}}+ \Sigma_{\hat{\boldsymbol{\psi}}}|\nonumber\\
    &=\log|\Sigma_{\hat{\boldsymbol{\psi}}}|+\log|\mathbf{I}+\mathbf{L}_{\mathbf{u}_n}^{-1}\mathbf{K}_{\mathbf{u}_n\mathbf{\hat{f}}}\Sigma_{\hat{\boldsymbol{\psi}}}^{-1}\mathbf{K}_{\mathbf{\hat{f}}\mathbf{u}_n}\mathbf{L}_{\mathbf{u}_n}^{-\top}|\nonumber\\
    &=(N_n+N_\mathcal{M})\log\sigma_\psi^2 + \log|\mathbf{D}_{\mathbf{u}_{n-1}}| \nonumber\\&\phantom{-}+\log|\mathbf{I}+\mathbf{L}_{\mathbf{u}_n}^{-1}\mathbf{K}_{\mathbf{u}_n\mathbf{\hat{f}}}\Sigma_{\hat{\boldsymbol{\psi}}}^{-1}\mathbf{K}_{\mathbf{\hat{f}}\mathbf{u}_n}\mathbf{L}_{\mathbf{u}_n}^{-\top}|,
\end{align}
where the inversion of covariance matrix is factorized as $\mathbf{K}_{\mathbf{u}_n\mathbf{u}_n}^{-1}=\mathbf{L}_{\mathbf{u}_n}^{-\top}\mathbf{L}_{\mathbf{u}_n}^{-1}$ using Cholesky decomposition to ensure computational stability. 
Define $\mathbf{D} = \mathbf{I}+\mathbf{L}_{\mathbf{u}_n}^{-1}\mathbf{K}_{\mathbf{u}_n\mathbf{\hat{f}}}\Sigma_{\hat{\boldsymbol{\psi}}}^{-1}\mathbf{K}_{\mathbf{\hat{f}}\mathbf{u}_n}\mathbf{L}_{\mathbf{u}_n}^{-\top}$ and using the Woodbury  matrix identity, we can calculate the inversion of $\mathbf{S}_\psi$:
\begin{align}\label{inv_Spsi}
    &\mathbf{S}_\psi^{-1} = (\mathbf{K}_{\mathbf{\hat{f}}\mathbf{u}_n}\mathbf{K}_{\mathbf{u}_n\mathbf{u}_n}^{-1}\mathbf{K}_{\mathbf{u}_n\mathbf{\hat{f}}}+ \Sigma_{\hat{\boldsymbol{\psi}}})^{-1} \nonumber\\
    &=\Sigma_{\hat{\boldsymbol{\psi}}}^{-1}-\Sigma_{\hat{\boldsymbol{\psi}}}^{-1} \mathbf{K}_{\mathbf{\hat{f}}\mathbf{u}_n} \mathbf{L}_{\mathbf{u}_n}^{-\top}\mathbf{D}^{-1}\mathbf{L}_{\mathbf{u}_n}^{-1}\mathbf{K}_{\mathbf{u}_n\mathbf{\hat{f}}}\Sigma_{\hat{\boldsymbol{\psi}}}^{-1}.
\end{align} 
Substituting the results in \eqref{log_det_Spsi} and \eqref{inv_Spsi} into \eqref{L_calculate}, we obtain
\begin{align}\label{eqn:L_hybrid_app}
&\mathcal{L}_\text{hybrid} \nonumber\\& =-\frac{1}{2}[(N_n+\mu_2N_\mathcal{M})\log(2\pi\sigma_\psi^2) + M\log(\mu_1) + N_\mathcal{M}\log(\mu_2)]\nonumber\\
&\quad + \frac{\mu_1}{2}\log\frac{|\mathbf{K}_{\mathbf{u}_{n-1}\mathbf{u}_{n-1}}|}{|\mathbf{S}_{\mathbf{u}_{n-1}}|} - \frac{1}{2}\log|\mathbf{D}|+ \frac{1}{2}\mathbf{c}^\top\mathbf{L}_{\mathbf{u}_n}^{-\top}\mathbf{D}^{-1}\mathbf{L}_{\mathbf{u}_n}^{-1}\mathbf{c} \nonumber\\
&\quad - \frac{1}{2\sigma_\psi^2}[\boldsymbol{\psi}_n^\top\boldsymbol{\psi}_n + \mu_2\boldsymbol{\psi}_\mathcal{M}^\top\boldsymbol{\psi}_\mathcal{M} + \text{tr}(\mathbf{Q_f}) + \mu_2\text{tr}(\mathbf{Q}_{\mathbf{f}_m})]\nonumber\\
&\quad - \frac{\mu_1}{2}[\text{tr}(\mathbf{D}_{\mathbf{u}_{n-1}}^{-1} \mathbf{Q}_{\mathbf{u}_{n-1}}) + \mathbf{m}_{\mathbf{u}_{n-1}}^\top\mathbf{S}_{\mathbf{u}_{n-1}}^{-1} \mathbf{m}_{\mathbf{u}_{n-1}}],
\end{align}
where $\mathbf{c} = \mathbf{K}_{\mathbf{u}_n\mathbf{\hat{f}}}\Sigma_{\hat{\boldsymbol{\psi}}}^{-1}\hat{\boldsymbol{\psi}} = \mathbf{K}_{\mathbf{u}_n\mathbf{u}_{n-1}}\mathbf{S}_{\mathbf{u}_{n-1}}^{-1}\mathbf{m}_{\mathbf{u}_{n-1}} +\frac{1}{\sigma_\psi^2}\mathbf{K}_{\mathbf{u}_n\mathbf{f}}\boldsymbol{\psi}$. 
Therefore, by computing \eqref{eqn:L_hybrid_app}, we can avoid direct operations on large matrices and instead work with a sequence of smaller matrices, which ensures a numerically stable implementation.

\subsection{Estimation of RSS Values for Non-measured Points}\label{Prediction}
As seen in \eqref{q opt(u)}, the optimal variational distribution has parameters $\mathbf{m}_{\mathbf{u}_n}=\mathbf{K}_{\mathbf{u}_n\mathbf{\hat{f}}}(\mathbf{K}_{\mathbf{\hat{f}}\mathbf{u}_n}\mathbf{K}_{\mathbf{u}_n\mathbf{u}_n}^{-1}\mathbf{K}_{\mathbf{u}_n\mathbf{\hat{f}}}+\Sigma_{\hat{\boldsymbol{\psi}}})^{-1}\hat{\boldsymbol{\psi}}$ and $\mathbf{S}_{\mathbf{u}_n}=\mathbf{K}_{\mathbf{u}_n\mathbf{u}_n}-\mathbf{K}_{\mathbf{u}_n\mathbf{\hat{f}}}(\mathbf{K}_{\mathbf{\hat{f}}\mathbf{u}_n}\mathbf{K}_{\mathbf{u}_n\mathbf{u}_n}^{-1}\mathbf{K}_{\mathbf{u}_n\mathbf{\hat{f}}}+\Sigma_{\hat{\boldsymbol{\psi}}})^{-1}\mathbf{K}_{\mathbf{\hat{f}}\mathbf{u}_n}$, which can be used for estimation at non-measured points as in \eqref{eq:prediction_3} and \eqref{eq:prediction_4}.
However, computing these parameters requires inverting a large matrix $\mathbf{K}_{\mathbf{\hat{f}}\mathbf{u}_n}\mathbf{K}_{\mathbf{u}_n\mathbf{u}_n}^{-1}\mathbf{K}_{\mathbf{u}_n\mathbf{\hat{f}}}+\Sigma_{\hat{\boldsymbol{\psi}}}$. 
To address this, we now provide the practical implementations of the estimation. 
We first rewrite the optimal variational distribution $q_{\text{opt}}^n(\mathbf{u}_n)$ in its natural parameter form:
\begin{align}
     q_{\text{opt}}^n(\mathbf{u}_n)&\propto  p(\mathbf{u}_n)\mathcal{N}(\hat{\boldsymbol{\psi}}; \mathbf{K}_{\mathbf{\hat{f}}\mathbf{u}_n}\mathbf{K}_{\mathbf{u}_n\mathbf{u}_n}^{-1}\mathbf{u}_n, \Sigma_{\hat{\boldsymbol{\psi}}}) \nonumber\\
    &= \mathcal{N}^{-1}\big(\mathbf{u}_n|\mathbf{m}_{\mathbf{u}_n}^\prime, \mathbf{S}_{\mathbf{u}_n}^\prime \big),
\end{align}
where $\mathbf{m}_{\mathbf{u}_n}^\prime$ and $\mathbf{S}_{\mathbf{u}_n}^\prime$ are obtained using the Woodbury matrix identity as
\begin{align}
    &\mathbf{S}_{\mathbf{u}_n}^\prime = \mathbf{S}_{\mathbf{u}_n}^{-1} = \mathbf{K}_{\mathbf{u}_n\mathbf{u}_n}^{-1} +     \mathbf{K}_{\mathbf{u}_n\mathbf{u}_n}^{-1}\mathbf{K}_{\mathbf{u}_n\mathbf{\hat{f}}}\Sigma_{\hat{\boldsymbol{\psi}}}^{-1}\mathbf{K}_{\mathbf{\hat{f}}\mathbf{u}_n}\mathbf{K}_{\mathbf{u}_n\mathbf{u}_n}^{-1},\nonumber\\
    &\mathbf{m}_{\mathbf{u}_n}^\prime = \mathbf{S}_{\mathbf{u}_n}^{-1}\mathbf{m}_{\mathbf{u}_n} = \mathbf{K}_{\mathbf{u}_n\mathbf{u}_n}^{-1}\mathbf{K}_{\mathbf{u}_n\mathbf{\hat{f}}}\Sigma_{\hat{\boldsymbol{\psi}}}^{-1}\hat{\boldsymbol{\psi}}.\nonumber
\end{align}
Then, by \eqref{eq:prediction_3} and \eqref{eq:prediction_4}, the predictive mean of RSS values at non-measured points $\mathbf{X}_n^*$ is 
\begin{align}\label{eqn:mu_app}
    &\boldsymbol{\mu}_n^* = \mathbf{K}_{\boldsymbol{\psi}^*{\mathbf{u}_n}}\mathbf{K}_{\mathbf{u}_n\mathbf{u}_n}^{-1}(\mathbf{S}_{\mathbf{u}_n}^{\prime})^{ -1}\mathbf{m}_{\mathbf{u}_n}^\prime \nonumber\\
    & = \mathbf{K}_{\boldsymbol{\psi}^*{\mathbf{u}_n}}\mathbf{K}_{\mathbf{u}_n\mathbf{u}_n}^{-1}(\mathbf{K}_{\mathbf{u}_n\mathbf{u}_n}^{-1} +     \mathbf{K}_{\mathbf{u}_n\mathbf{u}_n}^{-1}\mathbf{K}_{\mathbf{u}_n\mathbf{\hat{f}}}\Sigma_{\hat{\boldsymbol{\psi}}}^{-1}\mathbf{K}_{\mathbf{\hat{f}}\mathbf{u}_n}\mathbf{K}_{\mathbf{u}_n\mathbf{u}_n}^{-1})^{-1}\nonumber\\& \quad \times\mathbf{K}_{\mathbf{u}_n\mathbf{u}_n}^{-1}\mathbf{K}_{\mathbf{u}_n\mathbf{\hat{f}}}\Sigma_{\hat{\boldsymbol{\psi}}}^{-1}\hat{\boldsymbol{\psi}}\nonumber\\
    &=\mathbf{K}_{\boldsymbol{\psi}^*{\mathbf{u}_n}}\mathbf{L}_{\mathbf{u}_n}^{-\top}(\mathbf{I}+\mathbf{L}_{\mathbf{u}_n}^{-1}\mathbf{K}_{\mathbf{u}_n\mathbf{\hat{f}}}\Sigma_{\hat{\boldsymbol{\psi}}}^{-1}\mathbf{K}_{\mathbf{\hat{f}}\mathbf{u}_n}\mathbf{L}_{\mathbf{u}_n}^{-\top})^{-1}\mathbf{L}_{\mathbf{u}_n}^{-1}\mathbf{K}_{\mathbf{u}_n\mathbf{\hat{f}}}\Sigma_{\hat{\boldsymbol{\psi}}}^{-1}\hat{\boldsymbol{\psi}}\nonumber\\
    &=\mathbf{K}_{\boldsymbol{\psi}^*{\mathbf{u}_n}}\mathbf{L}_{\mathbf{u}_n}^{-\top}\mathbf{D}^{-1}\mathbf{L}_{\mathbf{u}_n}^{-1}\mathbf{K}_{\mathbf{u}_n\mathbf{\hat{f}}}\Sigma_{\hat{\boldsymbol{\psi}}}^{-1}\hat{\boldsymbol{\psi}},
\end{align}
and the predictive covariance is
\begin{align}\label{eqn:cov_app}
    &\mathbf{\Sigma}^*_n=\mathbf{K}_{\boldsymbol{\psi}^*\boldsymbol{\psi}^*}-\mathbf{K}_{\boldsymbol{\psi}^*{\mathbf{u}_n}}\mathbf{K}_{\mathbf{u}_n\mathbf{u}_n}^{-1}(\mathbf{K}_{\mathbf{u}_n\mathbf{u}_n}-(\mathbf{S}_{\mathbf{u}_n}^{\prime})^{ -1})\mathbf{K}_{\mathbf{u}_n\mathbf{u}_n}^{-1}\mathbf{K}_{{\mathbf{u}_n}\boldsymbol{\psi}^*}\nonumber\\
    &=\mathbf{K}_{\boldsymbol{\psi}^*\boldsymbol{\psi}^*}-\mathbf{K}_{\boldsymbol{\psi}^*{\mathbf{u}_n}}\mathbf{K}_{\mathbf{u}_n\mathbf{u}_n}^{-1}\mathbf{K}_{{\mathbf{u}_n}\boldsymbol{\psi}^*} + \mathbf{K}_{\boldsymbol{\psi}^*{\mathbf{u}_n}}\mathbf{K}_{\mathbf{u}_n\mathbf{u}_n}^{-1}\nonumber\\&\quad\times(\mathbf{K}_{\mathbf{u}_n\mathbf{u}_n}^{-1} +     \mathbf{K}_{\mathbf{u}_n\mathbf{u}_n}^{-1}\mathbf{K}_{\mathbf{u}_n\mathbf{\hat{f}}}\Sigma_{\hat{\boldsymbol{\psi}}}^{-1}\mathbf{K}_{\mathbf{\hat{f}}\mathbf{u}_n}\mathbf{K}_{\mathbf{u}_n\mathbf{u}_n}^{-1})\mathbf{K}_{\mathbf{u}_n\mathbf{u}_n}^{-1}\mathbf{K}_{{\mathbf{u}_n}\boldsymbol{\psi}^*}\nonumber\\
    &=\mathbf{K}_{\boldsymbol{\psi}^*\boldsymbol{\psi}^*}-\mathbf{K}_{\boldsymbol{\psi}^*{\mathbf{u}_n}}\mathbf{K}_{\mathbf{u}_n\mathbf{u}_n}^{-1}\mathbf{K}_{{\mathbf{u}_n}\boldsymbol{\psi}^*} + \mathbf{K}_{\boldsymbol{\psi}^*{\mathbf{u}_n}}\mathbf{L}_{\mathbf{u}_n}^{-\top}\nonumber\\&\quad \times (\mathbf{I}+\mathbf{L}_{\mathbf{u}_n}^{-1}\mathbf{K}_{\mathbf{u}_n\mathbf{\hat{f}}}\Sigma_{\hat{\boldsymbol{\psi}}}^{-1}\mathbf{K}_{\mathbf{\hat{f}}\mathbf{u}_n}\mathbf{L}_{\mathbf{u}_n}^{-\top})^{-1}\mathbf{L}_{\mathbf{u}_n}^{-\top}\mathbf{K}_{\mathbf{u}_n\boldsymbol{\psi}^*}\nonumber\\
    &=\mathbf{K}_{\boldsymbol{\psi}^*\boldsymbol{\psi}^*}-\mathbf{K}_{\boldsymbol{\psi}^*{\mathbf{u}_n}}\mathbf{K}_{\mathbf{u}_n\mathbf{u}_n}^{-1}\mathbf{K}_{{\mathbf{u}_n}\boldsymbol{\psi}^*}\nonumber\\
    &\phantom{-}+\mathbf{K}_{\boldsymbol{\psi}^*{\mathbf{u}_n}}\mathbf{L}_{\mathbf{u}_n}^{-\top}\mathbf{D}^{-1}\mathbf{L}_{\mathbf{u}_n}^{-\top}\mathbf{K}_{\mathbf{u}_n\boldsymbol{\psi}^*}.
\end{align}
Thus, by computing \eqref{eqn:mu_app} and \eqref{eqn:cov_app}, we avoid directly inverting large matrices and further apply the Cholesky decomposition of the covariance matrix $\mathbf{K}_{\mathbf{u}_n\mathbf{u}_n}=\mathbf{L}_{\mathbf{u}_n}\mathbf{L}_{\mathbf{u}_n}^\top$,which ensures numerically stable computations.

\end{document}